\documentclass[11pt,a4paper]{article}
\pdfoutput=1

\usepackage{jheppub}

\usepackage{amsmath, amssymb, amsthm} 
\usepackage{mathtools}
\usepackage{thmtools}
\usepackage{bm}
\usepackage{dsfont}
\usepackage{braket}
\usepackage{graphicx}

\numberwithin{equation}{section}

\usepackage{enumitem}
\setlist[itemize]{noitemsep}
\setlist[description]{noitemsep}

\usepackage[dvipsnames]{xcolor}
\usepackage{tikz}
\usepackage{pgfplots}
\usetikzlibrary{intersections,backgrounds}
\usepgfplotslibrary{fillbetween}
\pgfplotsset{compat = newest}

\usepackage[numbers, sort&compress]{natbib}
\usepackage{hyperref}
\hypersetup{colorlinks=true, linktoc=page, linkcolor=purple, citecolor=blue}

\newcommand{\negphantom}[1]{
    \ifmmode\settowidth{\dimen0}{$#1$}
    \else\settowidth{\dimen0}{#1}
    \fi
    \hspace*{-\dimen0}}
    
\makeatletter
\newcommand{\mask}[2]{{\mathpalette\mask@{{#1}{#2}}}}
\newcommand{\mask@}[2]{\mask@@{#1}#2}
\newcommand{\mask@@}[3]{%
  \settowidth{\dimen@}{$\m@th#1#2$}%
  \makebox[\dimen@]{$\m@th#1#3$}%
}
\makeatother
    
\newcommand{\sm}{\smallskip}

\usepackage{tikz}
\usetikzlibrary{decorations.pathmorphing,decorations.shapes}
\usetikzlibrary{decorations.pathreplacing,decorations.markings}
\usetikzlibrary{backgrounds}
\usetikzlibrary{positioning}
\usetikzlibrary{arrows}
\usetikzlibrary{shapes,shapes.geometric,shapes.misc}
\usepackage{pgfplots, pgfplotstable}
\pgfplotsset{width=10cm,compat=1.9}

\tikzstyle{tikzfig}=[baseline=-0.25em,scale=0.5]
\pgfkeys{/tikz/tikzit fill/.initial=0}
\pgfkeys{/tikz/tikzit draw/.initial=0}
\pgfkeys{/tikz/tikzit shape/.initial=0}
\pgfkeys{/tikz/tikzit category/.initial=0}

\pgfdeclarelayer{edgelayer}
\pgfdeclarelayer{nodelayer}
\pgfsetlayers{background,edgelayer,nodelayer,main}

\tikzstyle{none}=[inner sep=0mm]

\newcommand{\tikzfig}[1]{%
{\tikzstyle{every picture}=[tikzfig]
\IfFileExists{#1.tikz}
  {\input{#1.tikz}}
  {%
    \IfFileExists{./figures/#1.tikz}
      {\input{./figures/#1.tikz}}
      {\tikz[baseline=-0.5em]{\node[draw=red,font=\color{red},fill=red!10!white] {\textit{#1}};}}%
  }}%
}

\tikzset{->-/.style={decoration={
  markings,
  mark=at position #1 with {\arrow{>}}},postaction={decorate}}}

\tikzset{-<-/.style={decoration={
  markings,
  mark=at position #1 with {\arrow{<}}},postaction={decorate}}}

\tikzstyle{every loop}=[]
\begin{document}

\title{On the space of $2d$ integrable models}

\date{\today}

\author[a,b]{Lukas W. Lindwasser}
\affiliation[a]{
    Mani L. Bhaumik Institute for Theoretical Physics\\
    Department of Physics and Astronomy\\
    University of California, Los Angeles, CA 90095, USA}
\affiliation[b]{
    Department of Physics\\
    National Taiwan University\\
    Taipei 10617, Taiwan}
\emailAdd{lukaslindwasser@physics.ucla.edu}


\abstract{We study infinite dimensional Lie algebras, whose infinite dimensional mutually commuting subalgebras correspond with the symmetry algebra of $2d$ integrable models. These Lie algebras are defined by the set of infinitesimal, nonlinear, and higher derivative symmetry transformations present in theories with a left(right)-moving or (anti)-holomorphic current. We study a large class of such Lagrangian theories. We study the commuting subalgebras of the $2d$ free massless scalar, and find the symmetries of the known integrable models such as sine-Gordon, Liouville, Bullough–Dodd, and Korteweg–de Vries. Along the way, we find several new sequences of commuting charges, which we conjecture are charges of integrable models which are new deformations of a single scalar. After quantizing, the Lie algebra is deformed, and so are their commuting subalgebras.}


\maketitle

 \newpage

\section{Introduction}
\label{sec:intro}
Symmetries have played an essential role for physicists to not only organize our descriptions of the universe, but have also been an irreplaceable tool for computing observables from our theories. To calculate anything from some generic interacting quantum field theory, it is necessary to interpret the theory as some continuous deformation away from a highly symmetric theory where exact computations can be made, and calculate a perturbation series in the deformation parameter. One of the goals of a theoretical physicist is to find all highly symmetric/solvable points in the space of quantum field theories which may be used effectively by interpreting more realistic theories as deformations.

\sm

An interesting class of theories which are solvable are known as \textit{integrable} field theories, borrowing the term integrable from the Liouville–Arnold theorem \cite{Liouville1855}. An integrable Hamiltonian system $H(q_i(t),p_i(t),t)$ with $i=1,\dots,n$ has $n$ independent conserved charges $Q_i$ which mutually Poisson commute $\{Q_i,Q_j\}=0$. In this case, there exists a canonical transformation into action-angle coordinates $(I_i,\varphi_i)$ such that $I_i(t)=I_i(0)$ and $\varphi_i(t)=\omega_i(I)t+\varphi_i(0)$, making the whole system solvable by quadratures. To be integrable, a system must have as many commuting charges as there are degrees of freedom. 

\sm

A typical field theory in $d>1$ spacetime dimensions has infinitely many degrees of freedom, making the notion of integrability harder to define in this setting. Nevertheless there are several examples of integrable field theories, particularly when $d=2$, which have countably infinite mutually commuting charges, and in many of these cases explicit canonical transformations into action-angle coordinates have been found using the so called inverse scattering transform \cite{Gardner:1967wc,1972JETP...34...62Z,Ablowitz:1973fn,Ablowitz:1974ry}. 

\sm

Over the years there have been many techniques developed to demonstrate integrability, including amongst others B\"{a}cklund transformations, bi-Hamiltonian formulations, and Lax pairs/connections. Importantly, these techniques are not always guaranteed to work \cite{10.1063/1.529096}, and one requires further constraints on for instance 
 the Lax pairs to demonstrate integrability \cite{Sklyanin:1980ij,Maillet:1985ek,Maillet:1985ec}. There is not one definitive method for demonstrating integrability, and for each putative integrable model there requires a lot of ingenuity to do so. The common thread in these techniques however is in showing that a given model has countably infinite mutually commuting charges, and this is usually what is meant by integrable in the context of field theory.

\sm

Towards the goal of finding all solvable points in the space of field theories, we would like to understand the general structure of the space of integrable models. This problem is usually phrased in terms of what kinds of deformations of known integrable models preserve integrability. It has been shown for instance that for any $2d$ integrable quantum field theory there are infinitely many infinitesimal deformations which preserve infinitely many charges \cite{Smirnov:2016lqw}. It has furthermore been understood for a long time that there is a deep connection between $2d$ conformal field theories and $2d$ integrable models \cite{Sasaki:1987mm,Eguchi:1989hs,ZAMOLODCHIKOV1989641,Bazhanov:1994ft}. In this connection, one employs the Virasoro algebra, and searches for deformations of the conformal field theory which still preserves an infinite number of mutually commuting charges which are polynomials built out of the Virasoro generators $L_n$. In \cite{Sasaki:1987mm}, it was mentioned that ``...a very interesting problem (would be) to identify the huge group of symmetries which contain the above set of polynomials of the Virasoro generators as a commuting subalgebra". In this paper, we fully characterize the Lie algebra of symmetries offered by several classical Lagrangian $2d$ conformal field theories. This approach enables one to use Lie algebraic techniques to understand the space of $2d$ integrable models, by studying the infinite dimensional mutually commuting subalgebras of the Lie algebra of symmetries of a $2d$ conformal field theory, or more generally a theory with a left(right)-moving or (anti)-holomorphic current. As explained later in this paper, for each such subalgebra there is an associated infinite hierarchy of classically integrable models, and potentially more beyond those, which are deformations of the original theory.

\sm

To motivate this approach, consider the sine-Gordon model, which written in light-cone coordinates takes the form
\begin{align}
\label{eq:sineGordon}
    S = \int d^2x\Big(\partial_+\phi\partial_-\phi + \frac{m_0^2}{\beta^2}(\cos\beta\phi -1)\Big)    
\end{align}
This model is integrable \cite{Ablowitz:1973fn}, and may be thought of as a deformation of the single scalar conformal field theory, with a relevant deformation parameter $m_0^2$, and $\beta$ setting the periodicity $\phi\sim\phi+2\pi/\beta$ if $\phi$ is circle valued. This model has infinitely many mutually commuting charges, which generate infinitesimal symmetries of the action. The infinitesimal symmetry transformations split up into two types, taking the form $\delta_{2n+1\pm}\phi = \partial_{\pm}^{2n+1}\phi +\dots $, where the ellipsis includes nonlinear terms involving derivatives of $\phi$ \cite{Yoon:1975ma}. The first few non-trivial infinitesimal symmetries of the sine-Gordon model are
\begin{align}
    \delta_{1\pm}\phi &=\partial_\pm\phi \\
    \label{eq:singord3}
    \delta_{3\pm}\phi &= \partial_{\pm}^3\phi+\frac{1}{2}\beta^2(\partial_{\pm}\phi)^3 \\
    \delta_{5\pm}\phi &= \partial_\pm^5\phi + \frac{5}{2}\beta^2\big(\partial_\pm\phi(\partial_\pm^2\phi)^2+(\partial_\pm\phi)^2\partial_\pm^3\phi\big) + \frac{3}{8}\beta^4(\partial_\pm\phi)^5
\end{align}
The kinetic and potential terms in (\ref{eq:sineGordon}) are independently invariant under these transformations. The kinetic term, which on its own defines the single scalar conformal field theory, is invariant under a large space of symmetry transformations, which we will see later is best described as being generated by two copies of the space of all $1d$ Lagrangian functionals of a single field, the first copy involving the field $\partial_-\phi$ and the second copy involving the field $\partial_+\phi$. The potential term is not invariant under all of these transformations, but preserves an infinite set of mutually commuting infinitesimal symmetry transformations, which is a subset of the transformations the kinetic term is invariant under. The potential term may be interpreted as being built out of so-called screening charges of the single scalar conformal field theory which commute with the charges generating the symmetries of the sine-Gordon model. These screening charges may be used to define the left over mutually commuting symmetries \cite{Feigin:1991qm,Feigin:1993sb}.

\sm

If one was not yet aware of the sine-Gordon model and its infinitely many symmetries, they could \textit{discover} them by studying only the Lie algebra of symmetries of the single scalar conformal field theory, searching for its mutually commuting subalgebras. An integrable deformation that preserves the Poisson bracket structure $L=\partial_+\phi\partial_-\phi\to\partial_+\phi\partial_-\phi-V(\phi)$ and preserves those symmetries typically comes from a relatively simple additional calculation. Indeed, demanding that $\delta_{3\pm}V(\phi)$ be a total derivative already restricts $V(\phi)$ to be of the form \cite{doi:10.1098/rspa.1977.0012}
\begin{align}
    V(\phi) = Ae^{i\beta\phi} + Be^{-i\beta\phi} + C
\end{align}
From this perspective, rather than individually considering a putative integrable model, every infinite dimensional mutually commuting subalgebra defines integrable deformations of the single scalar conformal field theory.

\sm

In this paper, we study the Lie algebras of symmetries for a large class of $2d$ Lagrangian classical field theories which have a left(right)-moving or (anti)-holomorphic current, including the theory of a collection of free massless scalars, a collection of free massless fermions, and Wess–Zumino–Witten models. For each of these theories, a bootstrap program to search for integrable deformations may be initiated by finding all mutually commuting subalgebras. We study the mutually commuting subalgebras in detail in the case of a single scalar, finding some general properties of these subalgebras. We find that in order for two generators to commute, they must satisfy several differential equations, which we solve in many cases. Furthermore, all such subalgebras have a basis for which each independent generator has a different, nonzero order, to be defined later. We also find several mutually commuting subalgebras which, if infinite dimensional, define new infinite hierarchies of classically integrable models of a single scalar, along with finding subalgebras corresponding to all known integrable deformations of a single scalar. For some of these subalgebras, we study the putative integrable models they define and find exact wave solutions, providing some evidence of their integrability. 

\sm

Finally, we then consider how these Lie algebras, and correspondingly their mutually commuting subalgebras, are deformed after quantization. The Lie algebras themselves are deformed dramatically, and their subalgebras are under threat of disappearing. Nevertheless we find that for all examples we consider, the mutually commuting subalgebras are preserved to first order in $\alpha'$, the parameter which keeps track of quantum corrections.
\subsection{Conventions}
\label{sec:conv}
Apart from \autoref{sec:quantum}, we will work in Minkowski space, using light-cone coordinates with the convention
\begin{align}
    x^{\pm}=\frac{1}{\sqrt{2}}(x^0\pm x^1),\qquad \partial_{\pm}=\frac{1}{\sqrt{2}}(\partial_0\pm \partial_1)
\end{align}
\subsection{Outline}
\label{sec:outline}
In \autoref{sec:Euler}, we discuss some preliminaries, describing the general structure of the Lie algebras we will encounter in subsequent sections. In \autoref{sec:scalars} we describe the classical Lie algebras of symmetries present in the $2d$ field theories of a collection of free massless scalars, and do so for a collection of free massless fermions, and of Wess–Zumino–Witten models in \autoref{app:other}. Perhaps the most significant results are in \autoref{sec:deform}, where we study in detail the mutually commuting subalgebras of the Lie algebra in \autoref{sec:scalars} in the case $N=1$. Apart from some general properties of the mutually commuting subalgebras, we find several mutually commuting subalgebras, some of which we conjecture correspond to new integrable deformations of the single scalar. In \autoref{sec:quantum}, we study how the Lie algebra in \autoref{sec:deform} is deformed after quantization, and also find how the mutually commuting subalgebras are deformed to first order in $\alpha'$. Finally, in \autoref{sec:disc}, we discuss our results and present some questions for future research.
\section{Preliminaries} 
\label{sec:Euler}
In this section, we describe the general structure of the Lie algebras present in several $2d$ classical field theories possessing left-moving currents, which will be useful in finding integrable deformations of said theory. An essentially identical description can be presented when there are right-moving currents, which we omit. 

\sm

A current $j_{\mu}(x^+,x^-)$ satisfies the $2d$ conservation equation $\partial_- j_++\partial_+j_-=0$, written in light cone coordinates. In certain theories, typically conformal field theories, there exist left-moving currents, where $j_+(x^+,x^-)=0$. The conservation equation then implies $\partial_+j_-(x^+,x^-)=0$, or simply that $j_-(x^+,x^-)=j(x^-)$ does not depend on the right-moving coordinate $x^+$, after the equations of motion are imposed. Because of this we will refer to a left-moving current simply as $j(x^-)$ before the equations of motion are imposed unless stated otherwise. A standard example of a left-moving current is the energy-momentum tensor $T(x^-)=T_{--}(x^+,x^-)$ present in any $2d$ conformal field theory, due to the tracelessness of the energy-momentum tensor $T^{\mu}_{\;\;\mu} = T_{+-}=0$.

\sm

The existence of a left-moving current strongly constrains a theory, because it automatically implies an infinite number of additional left-moving currents. For instance, $j_v(x^-)=v(x^-)j(x^-)$ is left-moving for any left-moving function $v(x^-)$. It is this observation which is used to study $2d$ conformal field theories, organizing theories in terms of representations of the Witt algebra, or the Virasoro algebra when the theory is quantized, generated by currents of the form $(x^-)^{n+1}T(x^-)$. Most generally, any \textit{functional} $\ell(\{j(x^-),\partial_- j(x^-),\dots\},x^-)$ of $j(x^-)$ and its derivatives, which may be formally of arbitrary order, is left-moving. To avoid various technical subtleties, we will assume that the functionals considered here have a maximum finite order $k$ derivative of $j(x^-)$ which the functional depends on. We will use the shorthand $\ell(\{j_k(x^-)\},x^-)$ to indicate such a functional, where $j_k(x^-) = \partial_-^kj(x^-)$.

\sm

To consider the algebra generated by these currents, we must consider instead the conserved charges, i.e. the integrated currents
\begin{align}
    Q_{\ell} = \int dx^-\,\ell(\{j_k(x^-)\},x^-)
\end{align}
The Lie algebra of symmetries in turn is induced by the Poisson bracket of the theory
\begin{align}
\label{eq:QPoisson}
    \{Q_{\ell},Q_{\ell'}\} = Q_{\ell''}
\end{align}
Throughout this paper however, we will find it more convenient to describe the Lie algebra in terms of the infinitesimal symmetry transformation $\delta_\ell\mathcal{O}$ of some field $\mathcal{O}$, which is defined as usual through $Q_\ell$ and the Poisson bracket $\delta_\ell \mathcal{O}= \{Q_{\ell},\mathcal{O}\}$. The Lie algebra may then be expressed in terms of a commutator of infinitesimal transformations, which follows from the Jacobi identity of the underlying Poisson bracket
\begin{align}
\label{eq:ellcommutator}
   &\{Q_\ell,\{Q_{\ell'},\mathcal{O}\}\}-\{Q_{\ell'},\{Q_{\ell},\mathcal{O}\}\}=\{\{Q_\ell,Q_{\ell'}\},\mathcal{O}\} && \iff&&[\delta_\ell,\delta_{\ell'}]\mathcal{O} = \delta_{\ell''}\mathcal{O}
\end{align}
where $\ell''(\{j_{k''}(x^-)\},x^-)$ is defined via (\ref{eq:QPoisson}). This presentation of the Lie algebra is equivalent to (\ref{eq:QPoisson}) and in some sense more universal, as we have traded the theory dependent Poisson bracket structure with a commutator defined via composition of infinitesimal transformations. Once an infinitesimal symmetry transformation $\delta_\ell\mathcal{O}$ is identified, the Lie algebra follows immediately, as we will see through examples in \autoref{sec:scalars}, \autoref{sec:fermions}, and \autoref{sec:WZW}.

\sm

Assuming that $j(x^-)$ satisfies vanishing (at infinity) or periodic boundary conditions, unless it has a term that is independent of $j(x^-)$,\footnote{Such a term may lead to a central charge, which will be relevant in the quantum theory discussed in \autoref{sec:quantum}.} any functional which is a total derivative $\ell = \partial_- f(\{j_k(x^-)\},x^-)$ results in a vanishing charge. More generally, any such charge will at most depend on $j(x^-)$ and its derivatives on the spatial $x^-$ boundary, which will have a vanishing Poisson bracket with any local field. The set of functionals which generate the algebra then is the set of all functionals modulo total derivatives, or in other words the set of all $1d$ Lagrangians of $j(x^-)$. The Lie algebra defined by the commutator (\ref{eq:ellcommutator}) then acts on the vector space of $1d$ Lagrangians. Formulating the Lie algebra this way is quite natural, and we will see how the commutator $1d$ Lagrangian $\ell''(\{j_{k''}(x^-)\},x^-)$ depends on $\ell(\{j_k(x^-)\},x^-)$ and $\ell'(\{j_{k'}(x^-)\},x^-)$ only through their respective Euler–Lagrange equations in several examples. Furthermore, subalgebras may be defined in terms of the symmetries that a $1d$ Lagrangian possesses. In the presentation (\ref{eq:ellcommutator}), it is possible to forget about $Q_\ell$ and work directly with $\ell(\{j_k(x^-)\},x^-)$ and we refer to these as the generators.

\sm

We will say that two $1d$ Lagrangians $\ell$ and $\ell'$ are equivalent $\ell\sim \ell'$ if and only if they generate the same Euler–Lagrange equation, i.e. they differ by a total derivative $\ell=\ell' +\partial_-f$. Because of this equivalence relation, a $1d$ Lagrangian $\ell(\{j_k(x^-)\},x^-)$ which depends on $j(x^-)$ and its derivatives up to order $k$ may be equivalent to a Lagrangian $\ell'(\{j_{k'}(x^-)\},x^-)$ with $k'<k$. We will define the \textit{order of a Lagrangian} to be the minimum value of $k$ ($k$ being the maximum order derivative of $j(x^-)$ that $\ell$ depends on) that can be obtained by adding total derivatives. As a simple example, $\ell(\{j_4(x^-)\},x^-)=j(x^-)j_4(x^-)$ is order 2, because $jj_4\sim-j_1j_3\sim j_2j_2$. We will assume without loss of generality that $\ell(\{j_k(x^-)\},x^-)$ has been reduced such that $k$ represents the order of $\ell$.

\sm

A theory may possess more than one independent left-moving current, in the sense that they cannot be written as functionals of another left-moving current. If a theory possesses $N$ independent left-moving currents $j^a(x^-)$ $a=1,\dots,N$, then any functional involving all such currents $\ell(\{j^a_k(x^-)\},x^-)$ is a left-moving current.
\section{$2d$ free massless scalars}
\label{sec:scalars}
In this section, we review the Lie algebra of symmetries present in the classical theory of $N$ free massless scalars, emphasizing the presentation advocated for in \autoref{sec:Euler}. The formulae in this section will be particularly important for \autoref{sec:deform} and \autoref{sec:quantum}, where we will study the mutually commuting subalgebras of the single scalar in detail.

\sm

The classical theory of $N$ free massless real scalars $\phi^a$ has the action
\begin{align}
\label{eq:scalaraction}
    S = \int d^2x\, \partial_+\phi^a\partial_-\phi^a
\end{align}
It has $N$ independent left-moving currents $j^a(x^-)=\partial_-\phi^a$ (and $N$ independent right-moving currents). The left-moving energy-momentum tensor $T(x^-) = \partial_-\phi^a\partial_-\phi^a$ is not independent, as it is a functional of $j^a$, i.e. $T=j^aj^a$.

\sm

The Euler–Lagrange equation $\partial_+\partial_-\phi^a=0$ splits the phase space into left-moving and right-moving sectors. The fundamental Poisson bracket for the left-moving sector is
\begin{align}
    \{j^a(x^-),j^b(y^-)\}=-\frac{1}{2}\delta^{ab}\partial_{x^-}\delta(x^--y^-)
\end{align}
through which the Lie algebra of symmetries may be derived \cite{10.1063/1.1665772}. The resulting Lie algebra $\{Q_\ell,Q_{\ell'}\}=Q_{\ell''}$ will express $Q_{\ell''}$ in terms of functional derivatives of $Q_\ell$ and $Q_{\ell'}$ with respect to $j^a(x^-)$. Instead, we will describe in the next subsection the Lie algebra in terms of the infinitesimal transformation $\delta_\ell\phi^a$, encapsulated by equation (\ref{eq:scalarlie}). Much of the next two subsections are equivalent and can be translated to standard results in the language of Poisson brackets.
\subsection{Lie algebra}
\label{ssec:scalarlie}
The action (\ref{eq:scalaraction}) is invariant under the infinitesimal transformation
\begin{align}
\label{eq:scalarsym}
&\phi^a\to\phi^a+\epsilon\delta_{\ell}\phi^a\\
\label{eq:scalarsym1}
    &\delta_{\ell}\phi^a = \frac{1}{2}\mathcal{E}^a(\ell) = \frac{1}{2}\sum_{k=0}^{\infty}(-1)^k\partial_-^k\frac{\partial}{\partial j^a_k(x^-)}\ell(\{j^b_i(x^-)\},x^-)
\end{align}
for any functional $\ell(\{j^a_i(x^-)\},x^-)$, where $\mathcal{E}^a$ is a linear operator called the Euler operator for which $\mathcal{E}^a(\ell)$ gives the Euler–Lagrange equation for $j^a$, as derived from the $1d$ Lagrangian $\ell(\{j^a_i(x^-)\},x^-)$, and $\epsilon$ is some infinitesimal parameter. Since the Lagrangian is of finite order, the infinite sum in (\ref{eq:scalarsym}) will truncate at $k=i$. Under this transformation, the action becomes
\begin{align}
    \delta_\ell S &= \int d^2x\bigg(\partial_+\Big(\frac{1}{2}\mathcal{E}^a(\ell)\partial_-\phi^a-\ell(\{j^b_i(x^-)\},x^-)\Big) \nonumber \\
    &\hspace{1.5cm}+\partial_-\Big(\frac{1}{2}\mathcal{E}^a(\ell)\partial_+\phi^a+\sum_{k=0}^{\infty}\sum_{j=1}^{k}(-1)^{k-j}\partial_+\partial_-^{j}\phi^a\partial_-^{k-j}\frac{\partial}{\partial j^a_k(x^-)}\ell(\{j^b_i(x^-)\},x^-)\Big)\bigg) \nonumber\\
    & =0
\end{align}
and so the Noether current is $j_-=\ell(\{j^a_i(x^-)\},x^-)$ and $j_+=0$. 

\sm

In what follows it will be useful to introduce so called higher Euler operators $\mathcal{E}^a_n$ for any integer $n\geq 0$ \cite{10.1063/1.524104}, associated with the Euler operator $\mathcal{E}^a$ introduced in (\ref{eq:scalarsym}), which are defined in this context as
\begin{align}
    \mathcal{E}^a_n(\ell) = \sum_{k=0}^{\infty}(-1)^k{k+n \choose n}\partial_-^k\frac{\partial}{\partial j_{n+k}^a(x^-)}\ell(\{j^b_{i}(x^-)\},x^-)
\end{align}
Note that $\mathcal{E}^a_0=\mathcal{E}^a$ is the standard Euler operator. They have the important property that $\mathcal{E}^a_n(\partial_-\ell)=\mathcal{E}^a_{n-1}(\ell)$ for $n >0$, and when $n=0$, $\mathcal{E}^a(\partial_-\ell)=0$. With regard to the Euler operator, the converse is also true \cite{olver1993applications}, namely that if $\mathcal{E}^a(\ell)=0$ for any value of $a$, and $j^b(x^-)$ and its derivatives, then $\ell=\partial_-f$. The kernel of the Euler operator is therefore the space of functionals that are a total derivative. This sequence of operators has a product rule
\begin{align}
\label{eq:eulerprod}
    \mathcal{E}^a_n(fg)=\sum_{k=0}^{\infty}(-1)^k{k+n\choose n}\Big(\mathcal{E}^a_{n+k}(f)\partial_-^kg+\partial_-^kf\mathcal{E}^a_{n+k}(g)\Big)
\end{align}
and furthermore satisfy a composition rule
\begin{align}
\label{eq:eulercomp}
    \mathcal{E}^a_n\mathcal{E}^b_m = (-1)^n\sum_{k=0}^m{k+n \choose n}\frac{\partial}{\partial j^b_{n+k}(x^-)}\mathcal{E}^a_{m-k}
\end{align}
The symmetry transformations (\ref{eq:scalarsym}) associated with two Lagrangians $\ell$ and $\ell'$ do not in general commute. By direct computation, the commutator $[\delta_{\ell},\delta_{\ell'}]\phi^a$ is
\begin{align}
\label{eq:scalarliedirect}
    [\delta_{\ell},\delta_{\ell'}]\phi^a = \frac{1}{4}\sum_{k=0}^{\infty}\Big(\partial_-^{k+1}\mathcal{E}^b(\ell)\frac{\partial}{\partial j^b_k(x^-)}\mathcal{E}^a(\ell')-\partial_-^{k+1}\mathcal{E}^b(\ell')\frac{\partial}{\partial j^b_k(x^-)}\mathcal{E}^a(\ell)\Big)
\end{align}
Using both (\ref{eq:eulerprod}) and (\ref{eq:eulercomp}), we may write the right hand side of this equation in the form $\frac{1}{2}\mathcal{E}^a(\ell'')$, where $\ell''$ is a new Lagrangian
\begin{align}
\label{eq:scalarlie}
[\delta_{\ell},\delta_{\ell'}]\phi^a=\delta_{\ell''}\phi^a=\frac{1}{2}\mathcal{E}^a(\ell''),\qquad \ell''=\frac{1}{4}\Big(\partial_-\mathcal{E}^b(\ell)\mathcal{E}^b(\ell')-\partial_-\mathcal{E}^b(\ell')\mathcal{E}^b(\ell)\Big)
\end{align}
where $\ell''$ is written in a manifestly anti-symmetric in $(\ell\leftrightarrow \ell ')$ way, although because Lagrangians are defined modulo total derivatives, we may write $\ell ''=\frac{1}{2}\partial_-\mathcal{E}^b(\ell)\mathcal{E}^b(\ell')$. To show that (\ref{eq:scalarliedirect}) and (\ref{eq:scalarlie}) are equivalent, we compute $\frac{1}{2}\mathcal{E}^a(\ell'')$
\begin{align}
    \frac{1}{2}\mathcal{E}^a(\ell'')&=\frac{1}{4}\mathcal{E}^a\big(\partial_-\mathcal{E}^b(\ell)\mathcal{E}^b(\ell')\big) \nonumber\\
    &= \frac{1}{4}\sum_{k=0}^{\infty}(-1)^k\Big(\partial_-^{k+1}\mathcal{E}^b(\ell)\mathcal{E}^a_k\big(\mathcal{E}^b(\ell')\big)+\partial_-^{k}\mathcal{E}^b(\ell')\mathcal{E}^a_k\big(\partial_-\mathcal{E}^b(\ell)\big)\Big) \nonumber \\
    & =  \frac{1}{4}\Big(\sum_{k=0}^{\infty}(-1)^k\partial_-^{k+1}\mathcal{E}^b(\ell)\mathcal{E}^a_k\big(\mathcal{E}^b(\ell')\big)+\sum_{k=1}^{\infty}(-1)^k\partial_-^{k}\mathcal{E}^b(\ell')\mathcal{E}^a_{k-1}\big(\mathcal{E}^b(\ell)\big)\Big) \nonumber \\
    & = \frac{1}{4}\sum_{k=0}^{\infty}\Big(\partial_-^{k+1}\mathcal{E}^b(\ell)\frac{\partial}{\partial j^b_k(x^-)}\mathcal{E}^a(\ell')-\partial_-^{k+1}\mathcal{E}^b(\ell')\frac{\partial}{\partial j^b_{k}(x^-)}\mathcal{E}^a(\ell)\Big)
\end{align}
where the product rule (\ref{eq:eulerprod}) was used in the first line, $\mathcal{E}^a_k(\partial_-f)=\mathcal{E}^a_{k-1}(f)$ was used in the second line, and the composition rule (\ref{eq:eulercomp}) was used in the third line. This proves the equivalence of (\ref{eq:scalarliedirect}) and (\ref{eq:scalarlie}).

\sm

The commutator (\ref{eq:scalarlie}), which by construction is bilinear and satisfies the Jacobi identity, defines a Lie algebra on the vector space of $1d$ Lagrangians, which may be thought of as a nonlinear, higher derivative generalization of the Witt algebra of classical $2d$ infinitesimal conformal transformations of scalar fields. 
\subsection{Subalgebras}
\label{ssec:scalarsub}
The Lie algebra (\ref{eq:scalarlie}) of symmetry transformations of the action (\ref{eq:scalaraction}) described in \autoref{ssec:scalarlie} is infinite dimensional, with generators valued in the space of $1d$ Lagrangians of $j^a(x^-)$. To understand the structure of this Lie algebra, we describe several subalgebras.

\sm

If $\ell$ and $\ell'$ are polynomial functionals  of degree $n$ and $m$ respectively of $j^a$ and its derivatives, $\ell''$ in (\ref{eq:scalarlie}) will be at most a degree $n+m-2$ polynomial. The degree of $\ell''$ may be reduced if some terms in the polynomial are a total derivative, and hence equivalent to zero. Lagrangians $\ell$ of degree 2 and less, which generate linear transformations, therefore form a subalgebra. For the same reason, the space of polynomial Lagrangians forms a subalgebra.

\sm

This Lie algebra has at least an $n$ dimensional center, since the $1d$ Lagrangian $\ell^a=\partial_-\phi^a=j^a$ commutes with all other Lagrangians
\begin{align}
    \ell''^a=\frac{1}{2}\partial_-\mathcal{E}^b(\ell)\mathcal{E}^b(j^a)=\frac{1}{2}\partial_-\mathcal{E}^a(\ell)
\end{align}
Here we see that $\ell''^a$ is a total derivative, and so is equivalent to $\ell''^a=0$. Next, we would like to see what set of $1d$ Lagrangians $\ell(\{j_i^a(x^-)\},x^-)$ commute with the energy-momentum tensor $T=j^aj^a$. 
\begin{align}
    \frac{1}{2}\partial_-\mathcal{E}^b(T)\mathcal{E}^b(\ell)=\partial_-j^b\mathcal{E}^b(\ell)&=\sum_{k=0}^{\infty}(-1)^k\partial_-j^b\partial_-^k\frac{\partial}{\partial j_k^b(x^-)}\ell(\{j_i^a(x^-)\},x^-) \nonumber \\
    &\sim\sum_{k=0}^{\infty}\partial_-j_k^b\frac{\partial}{\partial j_k^b(x^-)}\ell(\{j_i^a(x^-)\},x^-) \nonumber \\
    & = \frac{d}{dx^-}\ell-\frac{\partial}{\partial x^-}\ell \nonumber \\
    &\sim -\frac{\partial}{\partial x^-}\ell(\{j_i^a(x^-)\},x^-)
\end{align}
where we have momentarily used the notation $\frac{d}{dx^-}$ to denote a total $x^-$ derivative and $\frac{\partial}{\partial x^-}$ to denote a partial $x^-$ derivative, varying $x^-$ while leaving $j_k^a(x^-)$ constant. We see in this case that in order for $\ell$ to commute with $T$, $\ell$ must have no explicit $x^-$ dependence, i.e. its Hamiltonian is conserved. The commutator of two such Lagrangians $\ell(\{j_i^a(x^-)\})$ and $\ell'(\{j_{i'}^a(x^-)\})$ has no explicit $x^-$ dependence either, and so the space of $1d$ Lagrangians with a conserved Hamiltonian forms a subalgebra.

\sm

The Witt algebra is a subalgebra, with Lagrangian generators $l_n=(x^-)^{n+1}T$. Indeed, the commutator between the two generators $l_n$ and $l_m$ is
\begin{align}
\label{eq:scalarWitt}
    \frac{1}{2}\partial_-\mathcal{E}(l_n)\mathcal{E}(l_m)&=2(n+1)(x^-)^{n+m+1}T + (x^-)^{n+m+2}\partial_-T\nonumber \\
    &\sim (n-m)(x^-)^{n+m+1}T=(n-m)l_{n+m}
\end{align}

\sm

If a $1d$ Lagrangian $\ell(\{j_i^a(x^-)\},x^-)$ is invariant under global $O(N)$ transformations $j^a\to G^{ab}j^b$ with $G\in O(N)$, then its Euler–Lagrange equation $\mathcal{E}^a(\ell)$ transforms covariantly $\mathcal{E}^a(\ell)\to G^{ab}\mathcal{E}^b(\ell)$. Looking at the form of the commutator (\ref{eq:scalarlie}), if both 
$\ell$ and $\ell'$ are $O(N)$ invariant, $\ell''$ is as well. The space of $O(N)$ invariant $1d$ Lagrangians therefore forms a subalgebra as well.

\sm

In the case of the subalgebra of Lagrangians with a conserved Hamiltonian, it was possible to define the subalgebra as the set of all Lagrangians which commute with the $2d$ energy-momentum tensor $T$. More generally, if the Lagrangian $\ell$ commutes with the Lagrangian $\ell'$, $\ell$ has a symmetry associated with $\ell'$ (and vice versa). In particular, $\ell$ is invariant under the infinitesimal transformation
\begin{align}
\label{eq:scalarcurrentsym}
    \delta_{\ell'}j^a =\frac{1}{2}\partial_-\mathcal{E}^a(\ell')
\end{align}
which is just the transformation (\ref{eq:scalarsym1}) acting on $j^a=\partial_-\phi^a$. Performing this infinitesimal transformation on $\ell(\{j_i^a(x^-)\},x^-)$, we get
\begin{align}
\label{eq:scalarlagrangeinv}
    \delta_{\ell'}\ell(\{j_i^a(x^-)\},x^-)&=\frac{1}{2}\sum_{k=0}^{\infty}\partial_-^{k+1}(\mathcal{E}^b(\ell'))\frac{\partial}{\partial j^b_k(x^-)}\ell(\{j_i^a(x^-)\},x^-) \nonumber \\
    &\sim \frac{1}{2}\sum_{k=0}^{\infty}(-1)^k\partial_-\mathcal{E}^b(\ell')\partial_-^{k}\frac{\partial}{\partial j^b_k(x^-)}\ell(\{j_i^a(x^-)\},x^-) \nonumber \\
    &=\frac{1}{2}\partial_-\mathcal{E}^b(\ell')\mathcal{E}^b(\ell)\sim 0
\end{align}
where the last line is equivalent to zero because $\ell$ and $\ell'$ commute in the sense of (\ref{eq:scalarlie}). This also follows directly from the Poisson bracket language $\delta_{\ell'}Q_\ell = \{Q_{\ell'},Q_\ell\}=\int dx^-\delta_{\ell'}\ell=0$.

\sm

The most interesting subalgebras for the purposes of this paper are the infinite dimensional mutually commuting subalgebras. Suppose there is a countably infinite set of Lagrangians $\ell_n$ with $n=1,2,\dots,\infty$ that satisfy
\begin{align}
\label{eq:integrablecommuting}
    \frac{1}{4}\Big(\partial_-\mathcal{E}^b(\ell_n)\mathcal{E}^b(\ell_m)-\partial_-\mathcal{E}^b(\ell_m)\mathcal{E}^b(\ell_n)\Big)\sim 0
\end{align}
for any $n,m$. These subalgebras are by definition possible symmetry algebras of $2d$ integrable models. Furthermore, the existence of such an infinite dimensional commuting subalgebra immediately implies the existence of infinitely many integrable deformations of the $N$ free massless scalar action (\ref{eq:scalaraction}). Indeed, the action
\begin{align}
\label{eq:scalardeform}
    S = \int d^2x\Big(\partial_+\phi^a\partial_-\phi^a +\lambda \, \ell_n(\{j_i^b\},x^-)\Big)
\end{align}
is invariant under the infinitesimal transformation $\delta_{\ell_m}\phi^a=\frac{1}{2}\mathcal{E}^a(\ell_m)$ for all $n,m$ and finite $\lambda$ by virtue of (\ref{eq:scalarlagrangeinv}) and (\ref{eq:integrablecommuting}). After deforming the theory as in (\ref{eq:scalardeform}), the currents associated with these symmetries will in general no longer be left-moving. Note that these actions are generically not Lorentz invariant, but will define equations of the Korteweg–de Vries type in terms of the fields $u^a(x^+,x^-)=\partial_-\phi^a(x^+,x^-)$. We will explore in detail the commuting subalgebras in the case of a single scalar in \autoref{sec:deform}.
\section{Integrable deformations of a single scalar}
\label{sec:deform}
In this section, we will study the mutually commuting subalgebras present in the Lie algebra described in \autoref{ssec:scalarlie}, in the case of a single scalar. If such a subalgebra is infinite dimensional, it automatically defines an infinite sequence of $2d$ integrable models of the Korteweg–de Vries type, as described at the end of \autoref{ssec:scalarsub}. In some special cases, the subalgebra defines the symmetry algebra of some Lorentz invariant $2d$ integrable model. We will restrict ourselves to consider $1d$ Lagrangians $\ell=\ell(\{j_k(x^-)\})$ which commute with the energy-momentum tensor $T=j^2$, so that the existence of a symmetry generated by $\ell$ implies a conserved charge $Q_\ell=\int dx^-\ell$.

\sm

Consider an order $n$ Lagrangian $\ell_n$ and an order $m$ Lagrangian $\ell_m$, with $m\geq n$. What conditions must these Lagrangians satisfy in order for them to commute? The commutator $\frac{1}{2}\partial_-\mathcal{E}(\ell_n)\mathcal{E}(\ell_m)$ must be a total derivative, or written as an equation, they must satisfy\footnote{Recall that $\mathcal{E}(\ell)=0$ for any value of $j(x^-)$ and its derivatives if and only if $\ell=\partial_-f$ \cite{olver1993applications}.}
\begin{align}
\label{eq:1scalcommnm}
    \mathcal{E}(\partial_-\mathcal{E}(\ell_n)\mathcal{E}(\ell_m))=0
\end{align}
Expanding out the Euler operators, this equation generates a large number of terms for any given $n,m$. For instance, when $(n,m)=(1,2)$ there are 880 terms, when $(n,m)=(1,3)$ there are $6,072$ terms, and when $(n,m)=(2,3)$ there are $54,845$ terms. The equation (\ref{eq:1scalcommnm}) depends on $j(x^-)$ and its derivatives up to order $2(n+m)$. Because we are looking for commuting $\ell_n$ and $\ell_m$ regardless of the value of $j(x^-)$ and its derivatives, we may regard this expression as a polynomial equation in $j_{m+1},\dots,j_{2(n+m)}$, whose coefficients on each monomial $j_{m+1}^{N_{m+1}}\cdots j_{2(n+m)}^{N_{2(n+m)}}$ must vanish independently. We will see shortly how making these coefficients vanish imposes differential equations that $\ell_n$ and $\ell_m$ must satisfy.

\sm

It is worth pausing here to reflect on what equation (\ref{eq:1scalcommnm}) achieves. The linear differential operator $\mathcal{E}$ acts on $\frac{1}{2}\partial_-\mathcal{E}(\ell_n)\mathcal{E}(\ell_m)$ and demands that it be a total derivative. This in turn imposes local differential equations on $\ell_n$ and $\ell_m$ as functions of $j_k(x^-)$ with $k=0,\dots,n$ and $k=0,\dots,m$, respectively. Looking for commuting Lagrangians then becomes a straightforward and systematic procedure.

\sm

Let us see how this works by proving a general property of two commuting $1d$ Lagrangians $\ell_n$ and $\ell_m$. In particular, we will prove that for $n,m>0$ and $m>n$, $\ell_n$ and $\ell_m$ can only commute if $\ell_m$ is quadratic in $j_m$. When $m=n$, the only $1d$ Lagrangian $\ell_n$ commutes with of the same order is a multiple of itself plus a $1d$ Lagrangian of lower order. The coefficient of the monomial $j_{2(n+m)}$ in the polynomial equation (\ref{eq:1scalcommnm}) can be obtained by noting that the Euler operator acting on an order $n>2$ Lagrangian $\ell_n$ has the expansion
\begin{align}
    \mathcal{E}(\ell_n)=(-1)^nj_{2n}\frac{\partial^2\ell_n}{\partial j_n^2}+n(-1)^nj_{2n-1}\partial_-\frac{\partial^2\ell_n}{\partial j_n^2}+\dots
\end{align}
and similarly for $\mathcal{E}(\ell_m)$. When $n=1,2$, there are modifications to this expansion because of the coincidences $n+1=2n,2n-1$, respectively. In these cases, we have instead
\begin{align}
    \mathcal{E}(\ell_1)&=-j_{2}\frac{\partial^2\ell_1}{\partial j_1^2}-j_{1}\frac{\partial^2\ell_1}{\partial j\partial j_1}+\frac{\partial\ell_1}{\partial j}\\
    \mathcal{E}(\ell_2)&=j_{4}\frac{\partial^2\ell_2}{\partial j_2^2}+j_{3}\Big(2\partial_-\frac{\partial^2\ell_2}{\partial j_2^2}-j_3\frac{\partial^3\ell_2}{\partial j_2^3}\Big)+\dots
\end{align}
Using this, and checking the cases when $n=1,2$ or $m=1,2$,     one obtains a uniform formula for $n,m>0$
\begin{align}
    &\mathcal{E}(\partial_-\mathcal{E}(\ell_n)\mathcal{E}(\ell_m))=\nonumber \\
    &=(-1)^{n+m}j_{2(n+m)}\Big((2m+1)\partial_-\frac{\partial^2\ell_n}{\partial j_n^2}\frac{\partial^2\ell_m}{\partial j_m^2} -(2n+1)\frac{\partial^2\ell_n}{\partial j_n^2}\partial_-\frac{\partial^2\ell_m}{\partial j_m^2}\Big)+\cdots
\end{align}
That this coefficient vanishes means that up to some normalization constant $C$
\begin{align}
     \frac{\partial^2\ell_m}{\partial j_m^2}=C\Big(\frac{\partial^2\ell_n}{\partial j_n^2}\Big)^{\frac{2m+1}{2n+1}}
\end{align}
From this we can deduce a few things. If $m> n$, $\ell_n$ does not depend on $j_m$, and we may freely integrate with respect to $j_m$ to get $\ell_m$
\begin{align}
\label{eq:ellm}
    \ell_m(j,\dots,j_m) = \frac{1}{2}C\Big(\frac{\partial^2\ell_n}{\partial j_n^2}\Big)^{\frac{2m+1}{2n+1}}j_m^2 + f_{m-1}(j,\dots,j_{m-1})
\end{align}
with $f_{m-1}(j,\dots,j_{m-1})$ some order $m-1$ functional yet to be determined. Here we have omitted a possible term linear in $j_m$, because such a term is always equivalent to a functional of order $m-1$ and can be absorbed into $f_{m-1}$. Indeed, the total derivative of some order $m-1$ functional $g_{m-1}(\{j_{m-1}(x^-)\})$ is
\begin{align}
    \partial_-g_{m-1}=j_m\frac{\partial g_{m-1}}{\partial j_{m-1}} + \sum_{k=0}^{m-2}j_{k+1}\frac{\partial g_{m-1}}{\partial j_k}\sim 0
\end{align}
and so $j_m\frac{\partial g_{m-1}}{\partial j_{m-1}}\sim -\sum_{k=0}^{m-2}j_{k+1}\frac{\partial g_{m-1}}{\partial j_k}$ is equivalent to a functional of order $m-1$. We see therefore that in order for $\ell_m$ to commute with $\ell_n$, $\ell_m$ can only depend on $j_m$ quadratically.

\sm

When $m=n$, we write $\ell_m=\ell_n'$ and find
\begin{align}
    \ell_n'(j,\dots,j_n)=C\ell_n(j,\dots,j_n) + f_{n-1}(j,\dots,j_{n-1})
\end{align}
Now, any Lagrangian commutes with itself, and so the first term in $\ell_n'$ trivially commutes with $\ell_n$. This means that up to such trivial terms, there are no two Lagrangians of the same order $n>0$ which commute! It follows from this that in any mutually commuting subalgebra of Lagrangians, there exists a basis of Lagrangians for which there is exactly one Lagrangian of a given order greater than 0. 

\sm

In the following subsections, we will consider solving the equation (\ref{eq:1scalcommnm}) for specific values of $n$ and $m$.
\subsection{$[\ell_0,\ell_0']=0$}
\label{ssec:ell00=0}
All order 0 Lagrangians without explicit $x^-$ dependence commute with each other. Indeed, the commutator between two such Lagrangians $\ell_0(j)$ and $\ell_0'(j)$ is 
\begin{align}
   \frac{1}{2}\partial_-\mathcal{E}(\ell_0)\mathcal{E}(\ell_0')= \frac{1}{2}j_1\frac{\partial^2\ell_0}{\partial j^2}\frac{\partial\ell_0'}{\partial j} =\frac{1}{2}\partial_-\Big(\int dj \frac{\partial^2\ell_0}{\partial j^2}\frac{\partial\ell_0'}{\partial j}\Big)\sim 0
\end{align}
\subsection{$[\ell_0,\ell_n]=0$}
\label{ssec:ell0n=0}
Considering what space of order $n$ Lagrangians $\ell_n$ commute with some order 0 Lagrangian $\ell_0$, the previous analysis in \autoref{sec:deform} does not apply. In this case, all order $n$ Lagrangians $\ell_n$ commute with the quadratic functional $\ell_0(j)=aj^2+bj+c$. Are there any restrictions on the space of $\ell_n$ when $\ell_0'''\neq 0$, where the prime now indicates a derivative with respect to $j$? 

\sm

When $n=1$, the commutator between $\ell_1(j,j_1)$ and $\ell_0(j)$ is a polynomial equation in $j_2$, and the coefficients of each monomial implies
\begin{align}
    &\frac{\partial}{\partial j}\Big(\ell_0'''\frac{\partial^2\ell_1}{\partial j_1^2}\Big)=0 
    &&\frac{\partial^3\ell_1}{\partial j_1^3}=-\frac{3}{j_1}\frac{\partial^2\ell_1}{\partial j_1^2}
\end{align}
which means that $\ell_1(j,j_1)$ takes the form
\begin{align}
\label{eq:ell01sol}
    \ell_1(j,j_1)=C\frac{1}{\ell_0'''(j)j_1} + f_0(j)
\end{align}
with $C$ a constant and $f_0(j)$ an arbitrary functional because of the analysis in \autoref{ssec:ell00=0}.

\sm

When $n=2$, the commutator is a polynomial equation in $j_3$ and $j_4$. In this case, the coefficient of the monomial $j_4$ implies
\begin{align}
    \frac{\partial^2}{\partial j_2^2}\Big(\ell_0'''\ell_2 - 3j_2\ell_0'''\frac{\partial\ell_2}{\partial j_2} - j_1^2\ell_0^{(4)}\frac{\partial\ell_2}{\partial j_2}-j_1\ell_0'''\frac{\partial\ell_2}{\partial j_1}\Big)=0
\end{align}
Here the order 2 Lagrangian $\ell_2(j,j_1,j_2)$ has infinitely many solutions of the form
\begin{align}
\ell_2(j,j_1,j_2)&=j_1g(u) + f_1(j,j_1) \\
\label{eq:uratioell0'''}
u&=\frac{j_2\ell_0'''(j)+j_1^2\ell_0^{(4)}(j)}{j_1^3\ell_0'''(j)}
\end{align}
where $g(u)$ is an arbitrary function, and $f_1(j,j_1)$ takes the form (\ref{eq:ell01sol}).

\sm

As we go higher in order, the solution space gets larger. For instance, when $n=3$, the commutator vanishing instead implies
\begin{align}
    \frac{\partial^2}{\partial j_3^2}\Big(j_1\ell_0'''\ell_3&-4j_3j_1\ell_0'''\frac{\partial\ell_3}{\partial j_3}-3j_2^2\ell_0'''\frac{\partial\ell_3}{\partial j_3}-6j_2j_1^2\ell_0^{(4)}\frac{\partial\ell_3}{\partial j_3} \nonumber \\
    & -j_1^4\ell_0^{(5)}\frac{\partial\ell_3}{\partial j_3} -3j_2j_1\ell_0'''\frac{\partial\ell_3}{\partial j_2}-j_1^3\ell_0^{(4)}\frac{\partial\ell_3}{\partial j_2} - j_1^2\ell_0'''\frac{\partial\ell_3}{\partial j_1}\Big)=0
\end{align}
The general solution to this equation is
\begin{align}
\ell_3(j,j_1,j_2,j_3)&=\,j_1g(u,v) +f_1(j,j_1) \\
 v&=\frac{(j_3j_1-3j_2^2)\ell_0'''^2-6j_2j_1^2\ell_0'''\ell_0^{(4)}-6j_1^4\ell_0^{(4)2}+j_1^4\ell_0'''\ell_0^{(5)}}{j_1^5\ell_0'''^2}
\end{align}
where $g(u,v)$ is an arbitrary function and $u$ is as in (\ref{eq:uratioell0'''}), $f_1(j,j_1)$ again takes the form (\ref{eq:ell01sol}). 

\sm

Important for the discussion in the following subsections, although there is a large space of order $n\geq 2$ Lagrangians which commute with some order 0 Lagrangian $\ell_0$ with $\ell_0'''\neq 0$, a generic order $n>0$ Lagrangian only commutes with $\ell_0$ when $\ell_0'''=0$. Because of the complexity of the equation (\ref{eq:1scalcommnm}), we will not continue this analysis to higher orders $n\geq 4$.
\subsection{$[\ell_1,\ell_2]=0$}
\label{ssec:ell12=0}
We now find the most general space of order 1 Lagrangians $\ell_1$ which commute with some order 2 Lagrangian $\ell_2$. Most well known $2d$ integrable models of a single scalar, including sine-Gordon, Liouville, Korteweg–de Vries, and modified Korteweg–de Vries, have commuting infinitesimal symmetry transformations generated by order 1 and 2 Lagrangians. In particular, sine-Gordon, Liouville and modified Korteweg–de Vries all have symmetries generated by the Lagrangians $j_1^2 + \alpha j^4$ and $j_2^2+ 10\alpha j^2j_1^2 + 2\alpha^2j^6$, while Korteweg–de Vries has symmetries generated by the Lagrangians $j_1^2 +2j^3$ and $j_2^2+10jj_1^2+5j^4$.

\sm

The equation (\ref{eq:1scalcommnm}) for $(n,m)=(1,2)$ generates many constraints on $\ell_1$ and $\ell_2$. We find that when $\ell_1(j,j_1)$ is not quadratic in $j_1$, in order for it to be a solution of (\ref{eq:1scalcommnm}) it must take the form
\begin{align}
    \ell_1(j,j_1) = \sqrt{\Big(\frac{cj_1}{\sqrt{a(j)}}+b(j)\Big)^2+a(j)} + d(j)
\end{align}
where $c$ is some constant, $d(j)$ is some quadratic functional of $j$, and $a(j)$ and $b(j)$ are functionals of $j$ satisfying
\begin{align}
    \frac{\partial^5}{\partial j^5}\Big(a(j)+b(j)^2\Big)=0
\end{align}
and $\ell_2(j,j_1,j_2)$ in turn is quadratic in $j_2$ of the form (\ref{eq:ellm}), where $f_1(j,j_1)$ has an explicit form in terms of $c$, $a(j)$, $b(j)$ and their derivatives written in (\ref{eq:1to2f1}) and (\ref{eq:p6}). 

\sm

When $\ell_1(j,j_1)$ is quadratic in $j_1$, we find a different set of solutions for $\ell_1(j,j_1)$, outlined in equations (\ref{eq:ell1quad}), (\ref{eq:eell1quad}), (\ref{eq:fell1quad}), and (\ref{eq:g1ansell1quad}). Restricting to polynomial $1d$ Lagrangians, we find only the order 1 $1d$ Lagrangians associated with the well known $2d$ integrable models of a single scalar mentioned above (up to an affine transformation $j\to a j+b$). The remainder of this subsection will be devoted to outlining each independent constraint coming from (\ref{eq:1scalcommnm}), which leads to these solutions.

\sm

To begin, $\ell_2$ has a higher order than $\ell_1$, and so using (\ref{eq:ellm}) we may write $\ell_2$ as
\begin{align}
    \ell_2(j,j_1,j_2)=\frac{1}{2}\Big(\frac{\partial^2\ell_1}{\partial j_1^2}\Big)^{5/3}j_2^2 + f_{1}(j,j_1)
\end{align}
setting $C=1$. Inputting this back into (\ref{eq:1scalcommnm}), we must now search for the order 1 functionals $\ell_1(j,j_1)$ and $f_1(j,j_1)$. We may now consider (\ref{eq:1scalcommnm}) as a polynomial equation in $j_2$, $j_3$, $j_4$, $j_5$, and $j_6$. Setting to zero the coefficient of each monomial $j_2^{N_1}j_3^{N_2}j_4^{N_3}j_5^{N_4}j_6^{N_5}$, one finds several equations the $\ell_1$ and $\ell_2$ must satisfy, many of which are redundant. Because of the complexity of the coefficients generated by (\ref{eq:1scalcommnm}), we will only summarize what each coefficient implies for $\ell_1(j,j_1)$ and $\ell_2(j,j_1,j_2)$. The coefficient of the monomial $j_2^3j_4$ imposes the following equation on $\ell_1(j,j_1)$
\begin{align}
\label{eq:1to2eqn1}
    40\Big(\frac{\partial^3\ell_1}{\partial j_1^3}\Big)^3-45\frac{\partial^2\ell_1}{\partial j_1^2}\frac{\partial^3\ell_1}{\partial j_1^3}\frac{\partial^4\ell_1}{\partial j_1^4} + 9\Big(\frac{\partial^2\ell_1}{\partial j_1^2}\Big)^2\frac{\partial^5\ell_1}{\partial j_1^5}=0
\end{align}
The general solution to this is
\begin{align}
\label{eq:1to2ABCD}
\ell_1(j,j_1)=\sqrt{A(j)j_1^2+B(j)j_1+C(j)}+D(j)
\end{align}
for any set of functionals $A(j)$, $B(j)$, $C(j)$ and $D(j)$. Here we have omitted a possible term linear in $j_1$, as this is a total derivative. A particularly important type of solution to (\ref{eq:1to2eqn1}) are those which satisfy $\frac{\partial^3\ell_1}{\partial j_1^3}=0$, i.e. $\ell_1(j,j_1)$ that are quadratic in $j_1$.\footnote{These can be obtained from (\ref{eq:1to2ABCD}) by taking appropriate infinite limits of $A,B,C,D$.} We will separately consider the cases when $\frac{\partial^3\ell_1}{\partial j_1^3}=0$ and $\frac{\partial^3\ell_1}{\partial j_1^3}\neq 0$.
\subsubsection{The case $\frac{\partial^3\ell_1}{\partial j_1^3}=0$}
If $\frac{\partial^3\ell_1}{\partial j_1^3}=0$, then $\ell_1$ is quadratic in $j_1$, and we may write it as
\begin{align}
\label{eq:ell1quad}
    \ell_1(j,j_1) = \frac{1}{2}e(j)j_1^2+f(j)
\end{align}
for some functionals $e(j)$, and $f(j)$. Inputting this back into (\ref{eq:1scalcommnm}), we now must search for $e(j)$, $f(j)$, and $f_1(j,j_1)$. The coefficient of the monomial $j_2j_4$ determines $f_1(j,j_1)$ to be of the form
\begin{align}
    f_1(j,j_1) = -\frac{5}{216}\frac{e'(j)^2+6e(j)e''(j)}{e(j)^{1/3}}j_1^4+\frac{1}{2}g_1(j)j_1^2+g_2(j)
\end{align}
where $g_1(j)$ and $g_2(j)$ are some functionals, and the prime here indicates a derivative with respect to $j$. We have also omitted another possible term linear in $j_1$. Inputting this again back into (\ref{eq:1scalcommnm}), we may now consider it a polynomial equation in $j_1$, $j_2$, $j_3$, and $j_4$. The coefficient of the monomial $j_1^3j_4$ imposes the following equation on $e(j)$
\begin{align}
    28e'(j)^3-36e(j)e'(j)e''(j)+9e(j)^2e'''(j)=0
\end{align}
The general solution to this is
\begin{align}
\label{eq:eell1quad}
    e(j)=\frac{1}{(aj^2+bj+c)^3}
\end{align}
with $a$, $b$, and $c$ constants. Next, the coefficients of the monomials $j_1j_4$ and $j_1^3j_2$ together impose equations on $g_1(j)$ and $f(j)$
\begin{align}
\label{eq:g1ell1quad}
    3e^3g_1'''-3e^2e'''g_1+30&ee'e''g_1-18e^2e''g_1'-12e^2e'g_1''+35ee'^2g_1'-35e'^3g_1=0\\
\label{eq:fell1quad}
    &f'''(j)=\frac{3}{5e(j)^{5/3}}(e(j)g_1'(j)-e'(j)g_1(j))
\end{align}
And finally, the coefficient of the monomial $j_1j_2$ imposes on $g_2(j)$ the equation
\begin{align}
\label{eq:g2ell1quad}
    g_2'''(j) = \frac{g_1(j)}{e(j)}f'''(j)
\end{align}

With $e(j)$ given by (\ref{eq:eell1quad}), equation (\ref{eq:g1ell1quad}) may be solved readily
\begin{align}
\label{eq:g1ansell1quad}
    g_1(j) = h_1e(j) + h_2\frac{j^2(bj+2c)^2}{(aj^2+bj+c)^5} + h_3\frac{j(bj+2c)(aj^2-c)}{(aj^2+bj+c)^5}
\end{align}
with $h_1$, $h_2$, and $h_3$ constants. $f(j)$ and $g_2(j)$ may then be found from (\ref{eq:fell1quad}) and (\ref{eq:g2ell1quad}), respectively.

\sm

What kind of polynomial order 1 Lagrangians can be obtained from $\ell_1(j,j_1)$ in (\ref{eq:ell1quad})? Recall that the well known $2d$ integrable models, including sine-Gordon, Liouville, Korteweg–de Vries and modified Korteweg–de Vries, have commuting infinitesimal symmetry transformations generated by polynomial order 1 and 2 Lagrangians, and so their respective order 1 Lagrangians must be found as a special case of (\ref{eq:ell1quad}). As reference, the order 1 Lagrangian associated with sine-Gordon, Liouville and modified Korteweg–de Vries is
\begin{align}
\label{eq:ellmkdv}
    \ell_{\text{mKdV}} = j_1^2 + \alpha j^4
\end{align}
with the value of $\alpha$ depending on the particular model, while the order 1 Lagrangian associated with Korteweg–de Vries is
\begin{align}
\label{eq:ellkdv}
    \ell_{\text{KdV}} = j_1^2 + 2j^3
\end{align}

\sm

To consider polynomial $\ell_1(j,j_1)$, we must make $e(j)$ a constant, which without loss of generality we may take $e(j)=2$. Equation (\ref{eq:g1ell1quad}) then implies $g_1'''(j)=0$, which means that $g_1(j)=g_1(0)+g_1'(0)j+\frac{1}{2}g_1''(0)j^2$ is a quadratic functional of $j$. Looking at equation (\ref{eq:fell1quad}), we see that $f(j)$ is therefore at most a degree 4 polynomial of $j$. The most general polynomial order 1 Lagrangian which commutes with an order 2 Lagrangian is therefore (up to a normalization)
\begin{align}
\label{eq:ell1poly}
    \ell_1(j,j_1)= j_1^2 + f(0) + f'(0)j + \frac{1}{2!}f''(0)j^2 + \frac{1}{3!}f'''(0)j^3 + \frac{1}{4!}f^{(4)}(0)j^4
\end{align}
Setting $f(0),f'(0),f''(0)=0$, the Lagrangians (\ref{eq:ellmkdv}) and (\ref{eq:ellkdv}) are obtained from this by setting $f'''(0)=0$ and $f^{(4)}(0)=0$, respectively. When both $f'''(0),f^{(4)}(0)\neq 0$, this $1d$ Lagrangian generates an infinitesimal symmetry transformation of what is sometimes referred as the Gardner equation, introduced in \cite{Kruskal1}.
\subsubsection{The case $\frac{\partial^3\ell_1}{\partial j_1^3}\neq 0$}
\label{ssec:1to2nonquad}
If $\frac{\partial^3\ell_1}{\partial j_1^3}\neq 0$, we return to the more general solution of (\ref{eq:1to2eqn1})
\begin{align}
\label{eq:1to2ABCD2}
\ell_1(j,j_1)=\sqrt{A(j)j_1^2+B(j)j_1+C(j)}+D(j)
\end{align}
Inputting this back into (\ref{eq:1scalcommnm}), the coefficient of the monomial $j_2^2j_4$ imposes
\begin{align}
\label{eq:1to2eqn2}
    9\Big(\frac{\partial^2\ell_1}{\partial j_1^2}\Big)^2\frac{\partial^5\ell_1}{\partial j \partial j_1^4}+40\Big(\frac{\partial^3\ell_1}{\partial j_1^3}\Big)^2\frac{\partial^3\ell_1}{\partial j \partial j_1^2}-21\frac{\partial^2\ell_1}{\partial j_1^2}\frac{\partial^4\ell_1}{\partial j_1^4}\frac{\partial^3\ell_1}{\partial j \partial j_1^2} - 24 \frac{\partial^2\ell_1}{\partial j_1^2}\frac{\partial^3\ell_1}{\partial j_1^3}\frac{\partial^4\ell_1}{\partial j \partial j_1^3}=0
\end{align}
which after plugging in (\ref{eq:1to2ABCD2}) implies
\begin{align}
    \frac{\partial}{\partial j}\big(B(j)^2-4A(j)C(j)\big)=0
\end{align}
we will therefore write $\ell_1(j,j_1)$ as
\begin{align}
\label{eq:ell1abdform}
\ell_1(j,j_1)=\sqrt{\Big(\frac{cj_1}{\sqrt{a(j)}}+b(j)\Big)^2+a(j)}+d(j)    
\end{align}
for some functionals $a(j)$, $b(j)$ and $d(j)$, and some constant $c\neq 0$. The coefficients of the monomials $j_2j_4$ and $j_4$ determine two different expressions for $f_1(j,j_1)$ in terms of $c$ and $a(j)$, $b(j)$, $d(j)$ and their derivatives. The two equations for $f_1(j,j_1)$ take the form
\begin{align}
\label{eq:1to2consis1}
    &\frac{\partial}{\partial j}h(j,j_1) = G_1(j,j_1) && \frac{\partial}{\partial j_1}h(j,j_1) = G_2(j,j_1)
\end{align}
where $h(j,j_1)=\frac{\partial^2}{\partial j_1^2}f_1/\frac{\partial^2}{\partial j_1^2}\ell_1$, and $G_1(j,j_1)$ and $G_2(j,j_1)$ are functions of $\ell_1(j,j_1)$ and its derivatives. Consistency between these two equations requires
\begin{align}
    \frac{\partial}{\partial j_1}G_1(j,j_1)=\frac{\partial}{\partial j}G_2(j,j_1)
\end{align}
Plugging our ansatz for $\ell_1(j,j_1)$ into this equation implies
\begin{align}
\label{eq:ell1dabeq}
    &d'''(j)=0
\end{align}
where the prime indicates a derivative with respect to $j$. $d'''(j)=0$ means that $d(j)=d(0)+d'(0)j+\frac{1}{2}d''(0)j^2$ is a quadratic functional of $j$, which commutes with all $1d$ Lagrangians $\ell(\{j_i(x^-)\})$ with no explicit $x^-$ dependence. $f_1(j,j_1)$ in turn is found to take the form
\begin{align}
\label{eq:1to2f1}
    f_1(j,j_1)=\frac{1}{2}\Big(\frac{\partial^2\ell_1}{\partial j_1^2}\Big)^{5/3}p_6(j,j_1) + e\ell_1(j,j_1) + g(j)
\end{align}
with $e$ some constant and $g(j)$ some functional of $j$. $p_6(j,j_1)$ is a degree six polynomial in $j_1$, whose coefficients depend on $a(j)$, $b(j)$ and their derivatives
\begin{align}
\label{eq:p6}
    &p_6(j,j_1)=c^2j_1^6\Big(\frac{2bb''}{3a^3}+\frac{4a''}{3a^3}+\frac{5b'^2}{3a^3}+\frac{2a''b^2}{3a^4}+\frac{2a'bb'}{3a^4}-\frac{5a'^2}{6a^4}-\frac{a'^2b^2}{a^5}\Big) \nonumber\\
    &+cj_1^5\Big(-\frac{2b''}{a^{3/2}}+\frac{a'b'}{a^{5/2}}+\frac{26bb'^2}{3a^{5/2}}+\frac{19a''b}{3a^{5/2}}+\frac{8b^2b''}{3a^{5/2}}+\frac{10a'b^2b'}{3a^{7/2}}+\frac{10a''b^3}{3a^{7/2}}-\frac{25a'^2b}{6a^{7/2}}-\frac{5a'^2b^3}{a^{9/2}}\Big) \nonumber \\
    &+5j_1^4\Big(\frac{a''}{3a}+\frac{2b'^2}{3a}-\frac{4bb''}{3a} +\frac{8a''b^2}{3a^2}+\frac{2b^3b''}{3    a^2}+\frac{11b^2b'^2}{3a^2}\nonumber \\
    &\hspace{1.5cm}-\frac{5a'^2}{12a^2}+\frac{4a'bb'}{3a^2}+\frac{4a''b^4}{3a^3}+\frac{4a'b^3b'}{3a^3}-\frac{13a'^2b^2}{6a^3}-\frac{2a'^2b^4}{a^4}\Big) \nonumber \\
    &-\frac{5j_1^3}{c}\Big(\frac{2}{3}a^{1/2}b''+\frac{2b^2b''}{a^{1/2}}-\frac{a'b'}{a^{1/2}}-\frac{2bb'^2}{a^{1/2}}-\frac{a''b}{a^{1/2}}+\frac{a'^2b}{a^{3/2}}-\frac{3a'b^2b'}{a^{3/2}}\nonumber \\
    &\hspace{1.5cm}-\frac{3a''b^3}{a^{3/2}}-\frac{4b^3b'^2}{a^{3/2}}  +\frac{19a'^2b^3}{6a^{5/2}}-\frac{4a'b^4b'}{3a^{5/2}}-\frac{4a''b^5}{3a^{5/2}}+\frac{2a'^2b^5}{a^{7/2}}\Big) \nonumber \\
    &-\frac{5j_1^2}{c^2}\Big(\frac{4}{3}abb''-2a'bb'-a''b^2-2b^2b'^2+2b^3b''+\frac{5a'^2b^2}{4a}-\frac{3a'b^3b'}{a}\nonumber \\
    &\hspace{1.5cm}-\frac{7b^4b'^2}{3a}-\frac{5a''b^4}{3a}+\frac{2b^5b''}{3a}+\frac{7a'^2b^4}{3a^2}-\frac{2a'b^5b'}{3a^2}-\frac{2a''b^6}{3a^2}+\frac{a'^2b^6}{a^3}\Big) \nonumber \\
    &-\frac{j_1}{c^3}(a+b^2)\Big(\frac{4}{3}a^{3/2}b''-2a^{1/2}a'b'+4a^{1/2}b^2b''+\frac{a'^2b}{2a^{1/2}}-\frac{6a'b^2b'}{a^{1/2}} \nonumber \\
    &\hspace{2.5cm}-\frac{10b^3b'^2}{3a^{1/2}}-\frac{2a''b^3}{3a^{1/2}}+\frac{8b^4b''}{3a^{1/2}}+\frac{7a'^2b^3}{3a^{3/2}}-\frac{2a'b^4b'}{3a^{3/2}}-\frac{2a''b^5}{3a^{3/2}}+\frac{a'^2b^5}{a^{5/2}} \Big)\nonumber \\
    &-\frac{1}{12c^4}(a+b^2)^2\Big(4aa''-3a'^2+8ab'^2+4a''b^2-12a'bb'+8abb''+8b^3b''-4b^2b'^2\Big)
\end{align}
The last constraint comes from the coefficient of the monomial $j_2$, which requires that
\begin{align}
\label{eq:ell1beq}
    \frac{\partial^5}{\partial j^5}\Big(a(j)+b(j)^2\Big)=0
\end{align}
and so $f(j)=a(j)+b(j)^2$ is at most a degree 4 polynomial of $j$. With no more constraints on $\ell_1$, (\ref{eq:ell1abdform}), (\ref{eq:ell1dabeq}), and (\ref{eq:ell1beq}) define the full parameter space of solutions for $\ell_1$, with free parameters we may take as $a(0)$, $a'(0)$, $a''(0)$, $a'''(0)$, $a^{(4)}(0)$, $c$, $d(0)$, $d'(0)$, $d''(0)$, and $b(j)$ undetermined. Finally, in order for $\ell_2$ to commute with $\ell_1$, $g(j)$ must commute with $\ell_1$ independently from the other terms in $\ell_2$, which forces $g(j)=g(0)+g'(0)j+\frac{1}{2}g''(0)j^2$ to be a quadratic functional of $j$ because of the analysis in \autoref{ssec:ell0n=0}, just like $d(j)$.
\subsection{$[\ell_2,\ell_3]=0$}
\label{ssec:ell23=0}
We now find the space of order 2 Lagrangians $\ell_2$ which commute with some order $3$ Lagrangian $\ell_3$. The $2d$ integrable models mentioned in the beginning of \autoref{ssec:ell12=0} also have an infinitesimal symmetry transformation generated by an order 3 Lagrangian (as well as Lagrangians of every order) that commutes with both the order 1 and 2 Lagrangians of their respective theories. Meanwhile, the Bullough–Dodd model does not have a symmetry generated by an order 1 Lagrangian, but does have commuting infinitesimal symmetry transformations generated by order 2 and 3 Lagrangians. In particular, the order 2 Lagrangian associated with the Bullough–Dodd model is $j_2^2-5j_1^3+45j^2j_1^2+27j^6$, while the order 3 Lagrangian is $j_3^2-21j_1j_2^2+63j^2j_2^2-21j_1^4-126j^2j_1^3+1134j^4j_1^2+243j^8$.

\sm

The equation (\ref{eq:1scalcommnm}) for $(n,m)=(2,3)$ generates many constraints on $\ell_2$ and $\ell_3$. We find that when $\ell_2(j,j_1,j_2)$ is not quadratic in $j_2$, in order for it to be a solution of (\ref{eq:1scalcommnm}) it must take the form
\begin{align}
    \ell_2(j,j_1,j_2) = \big(aj_2+b(j,j_1)\big)^{1/3} + c(j)
\end{align}
where $a$ is some constant, $c(j)$ is some quadratic functional of $j$, and $b(j,j_1)$ takes the form
\begin{align}
b(j,j_1)=a^{3/2}A(j)j_1^3+aB(j)j_1^2+a^{1/2}C(j)j_1+D(j)
\end{align}
where $A(j)$, $B(j)$, $C(j)$, and $D(j)$ are functionals of $j$ that satisfy complicated constraints outlined in equations (\ref{eq:Bredef}), (\ref{eq:DE1}), (\ref{eq:DE2}), and (\ref{eq:A}). $\ell_3(j,j_1,j_2,j_3)$ is quadratic in $j_3$ of the form (\ref{eq:ellm}), where $f_2(j,j_1,j_2)$ has an explicit form written in (\ref{eq:f2ell2to3}) and (\ref{eq:p3}). Because $\ell_2(j,j_1,j_2)$ is not quadratic in $j_2$, it cannot commute with an order 1 Lagrangian.

\sm

When $\ell_2(j,j_1,j_2)$ is quadratic in $j_2$, we find that all such order 2 Lagrangians which can possibly commute with an order 1 Lagrangian, do. This will be made more precise shortly in \autoref{ssec:ell2quad}. Restricting to polynomial $1d$ Lagrangians, we find only the order 2 $1d$ Lagrangians associated with the well known $2d$ integrable models of a single scalar mentioned above. The remainder of this subsection will be devoted to outlining each independent constraint coming from (\ref{eq:1scalcommnm}), which leads to these solutions.

\sm

To begin, because of (\ref{eq:ellm}), $\ell_3$ takes the form
\begin{align}
    \ell_3(j,j_1,j_2,j_3)=\frac{1}{2}\Big(\frac{\partial^2\ell_2}{\partial j_2^2}\Big)^{7/5}j_3^2+f_2(j,j_1,j_2)
\end{align}
setting $C=1$. Inputting this back into (\ref{eq:1scalcommnm}), we must now search for the order 2 functionals $\ell_2(j,j_1,j_2)$ and $f_2(j,j_1,j_2)$. We may now consider (\ref{eq:1scalcommnm}) a polynomial equation in $j_3$, $j_4$, $j_5$, $j_6$, $j_7$, and $j_8$. Setting to zero the coefficient of each monomial $j_3^{N_1}j_4^{N_2}j_5^{N_3}j_6^{N_4}j_7^{N_5}j_8^{N_6}$, one finds several equations the $\ell_2$ and $\ell_3$ must satisfy. Again, because of the complexity of the coefficients, we will only summarize what each coefficient implies for $\ell_2(j,j_1,j_2)$ and $\ell_3(j,j_1,j_2,j_3)$. The first constraint comes from the coefficient of the monomial $j_4j_8$, which imposes the following equation on $\ell_2(j,j_1,j_2)$
\begin{align}
\label{eq:j8j4eq}
    8\frac{\partial^3\ell_2}{\partial j_2^3}\partial_-\frac{\partial^2\ell_2}{\partial j_2^2}-5\frac{\partial^2\ell_2}{\partial j_2^2}\partial_-\frac{\partial^3\ell_2}{\partial j_2^3}=0
\end{align}
We will separately consider the cases when $\frac{\partial^3\ell_2}{\partial j_2^3}=0$ and $\frac{\partial^3\ell_2}{\partial j_2^3}\neq 0$.
\subsubsection{The case $\frac{\partial^3\ell_2}{\partial j_2^3}=0$}
\label{ssec:ell2quad}
If $\frac{\partial^3\ell_2}{\partial j_2^3}=0$, then $\ell_2$ is quadratic in $j_2$, and because of the preceding discussion in \autoref{sec:deform} it has a chance of commuting with some order 1 Lagrangian, which we denote as $\ell_1$. In this case then, we will restrict for the moment to finding order 2 Lagrangians which commute with an order 3 Lagrangian, and have the possibility of commuting with an order 1 Lagrangian. We will find that all such order 2 Lagrangians commute with both an order 1 and 3 Lagrangian. We may without loss of generality write $\ell_2$ suggestively as
\begin{align}
    \ell_2(j,j_1,j_2)=\frac{1}{2}\Big(\frac{\partial^2}{\partial j_1^2}\ell_1(j,j_1)\Big)^{5/3}j_2^2 + f_{1}(j,j_1)
\end{align}
identifying $\ell_1(j,j_1)$ as the potential order 1 Lagrangian that $\ell_2$ commutes with, and $f_1(j,j_1)$ along with $f_2(j,j_1,j_2)$ are to be determined. Inputting this back into (\ref{eq:1scalcommnm}), the coefficient of the monomial $j_3j_8$ determines $f_2(j,j_1,j_2)$ in terms of $\ell_1(j,j_1)$
\begin{align}
    f_2(j,j_1,j_2) =& -\frac{7}{24}j_2^4\Big(\frac{\partial^2\ell_1}{\partial j_1^2}\Big)^{4/3}\frac{\partial^4\ell_1}{\partial j_1^4}-\frac{7}{6}j_2^3\Big(\frac{\partial^2\ell_1}{\partial j_1^2}\Big)^{4/3}\Big(\frac{\partial^3\ell_1}{\partial j\partial j_1^2}+j_1\frac{\partial^4\ell_1}{\partial j\partial j_1^3}\Big) \nonumber \\
    &+\frac{1}{2}g(j,j_1)j_2^2+h(j,j_1)
\end{align}
with $g(j,j_1)$ and $h(j,j_1)$ left undetermined, and we have omitted a possible term linear in $j_2$. Now that the explicit $j_2$ dependence has been determined, the equation (\ref{eq:1scalcommnm}) may now be considered a polynomial in $j_2$, $j_3$, $j_4$, $j_5$, $j_6$, $j_7$, and $j_8$. Next, the coefficients of $j_2j_3^2j_6$ and $j_3^2j_6$ impose the equations (\ref{eq:1to2eqn1}) and (\ref{eq:1to2eqn2}) on $\ell_1(j,j_1)$, respectively. Hence, we may write $\ell_1(j,j_1)$ as (\ref{eq:ell1abdform})
\begin{align}
\ell_1(j,j_1)=\sqrt{\Big(\frac{cj_1}{\sqrt{a(j)}}+b(j)\Big)^2+a(j)}+d(j)    
\end{align}
The coefficient of the monomial $j_8$ determines $g(j,j_1)$ in terms of $\ell_1(j,j_1)$ and $f_1(j,j_1)$,
\begin{align}
    g(j,j_1) = \frac{7}{5}\Big(\frac{\partial^2\ell_1}{\partial j_1^2}\Big)^{2/3}\frac{\partial^2f_1}{\partial j_1^2} + \frac{7}{9}j_1^2\Big(\frac{\partial^2\ell_1}{\partial j_1^2}\Big)^{1/3}\Big(\Big(\frac{\partial^3\ell_1}{\partial j\partial j_1^2}\Big)^2-3\frac{\partial^2\ell_1}{\partial j_1^2}\frac{\partial^4\ell_1}{\partial j^2\partial j_1^2}\Big)+e\Big(\frac{\partial^2\ell_1}{\partial j_1^2}\Big)^{5/3}
\end{align}
with $e$ some constant. Because of the analysis in \autoref{ssec:ell12=0}, in order for $\ell_2$ to commute with an order 1 Lagrangian $\ell_1(j,j_1)$, $f_1(j,j_1)$ must equal (\ref{eq:1to2f1}). We will now restrict to taking $f_1(j,j_1)$ to be (\ref{eq:1to2f1}) with $a(j)$ and $b(j)$ and $c$ still arbitrary, so that $\ell_2(j,j_1,j_2)$ has a chance of commuting with an order 1 Lagrangian $\ell_1(j,j_1)$. By doing this, we lose solutions of $\ell_2(j,j_1,j_2)$ which do not commute with an order 1 Lagrangian, but have not yet ruled out all such $\ell_2(j,j_1,j_2)$. With $f_1(j,j_1)$ given, this determines $g(j,j_1)$ completely in terms of $a(j)$, $b(j)$, and $c$.

\sm

Finally, the coefficients of the monomials $j_2j_6$ and $j_6$ determine two different expressions for $h(j,j_1)$ in terms of $a(j)$, $b(j)$, $c$ and their derivatives. In a similar fashion to the discussion around equations (\ref{eq:1to2consis1})–(\ref{eq:ell1dabeq}), for the two expressions to be equal, consistency demands
\begin{align}
    \frac{\partial^5}{\partial j^5}\Big(a(j)+b(j)^2\Big)=0
\end{align}
This completely determines $\ell_1(j,j_1)$ up to $d(j)$, which we may take to be a quadratic polynomial as in \autoref{ssec:1to2nonquad}. This is precisely the same $\ell_1(j,j_1)$ determined in \autoref{ssec:1to2nonquad}. We therefore find that all quadratic order 2 Lagrangians $\ell_2(j,j_1,j_2)$ which commute with an order $3$ Lagrangian $\ell_3(j,j_1,j_2,j_3)$, and furthermore have the \textit{possibility} of commuting with an order 1 Lagrangian $\ell_1(j,j_1)$, do commute with $\ell_1(j,j_1)$.

\sm

Let us now restrict to searching for polynomial order 2 Lagrangians which commute with a polynomial order 3 Lagrangian. In this case, we expect to find the order 2 and 3 Lagrangians associated with sine-Gordon, Liouville, modified Korteweg–de Vries, and Korteweg–de Vries models, but also the Bullough–Dodd model, which does not have a symmetry generated by an order 1 Lagrangian. We may write the order 2 and order 3 Lagrangians as
\begin{align}
    \ell_2(j,j_1,j_2) &= j_2^2 + f_1(j,j_1) \\
    \ell_3(j,j_1,j_2,j_3)& =j_3^2 + f_2(j,j_1,j_2)
\end{align}
Inputting this into equation (\ref{eq:1scalcommnm}), the equation may be considered a polynomial in $j_3$, $j_4$, $j_5$, $j_6$, $j_7$, and $j_8$. The coefficient of the monomial $j_8$ imposes the equation
\begin{align}
    \partial_-\frac{\partial^2f_2}{\partial j_2^2}= \frac{7}{5}\partial_-\frac{\partial^2f_1}{\partial j_1^2}
\end{align}
And so $f_2(j,j_1,j_2)$ takes the form
\begin{align}
    f_2(j,j_1,j_2) = \frac{1}{2}\Big(\frac{7}{5}\frac{\partial^2f_1}{\partial j_1^2} + A\Big)j_2^2 + B(j,j_1)
\end{align}
where $A$ is some constant, and $B(j,j_1)$ is some functional, and we have omitted a possible term linear in $j_2$. Inputting this back into (\ref{eq:1scalcommnm}), we may now consider it a polynomial equation in $j_2$, $j_3$, $j_4$, $j_5$, and $j_6$. The coefficient of the monomial $j_3j_6$ implies
\begin{align}
    \partial_-\frac{\partial^3f_1}{\partial j_1^3}=0
\end{align}
And so $f_1(j,j_1)$ is at most a cubic polynomial, with a constant coefficient on $j_1^3$
\begin{align}
    f_1(j,j_1) = \frac{1}{6}aj_1^3 + \frac{1}{2}C(j)j_1^2 + D(j)
\end{align}
for some constant $a$, some functionals $C(j)$, and $D(j)$, and we have omitted a possible term linear in $j_1$. Next, the coefficient of the monomial $j_6$ determines $B(j,j_1)$
\begin{align}
    B(j,j_1) =&\, \frac{7}{600}\big(a^2-15C''(j)\big)j_1^4 + \frac{1}{300}a\big(25A+14C(j)\big)j_1^3 \nonumber \\
    &+ \frac{1}{100}\big(-70d+25AC(j)+7C(j)^2+70D''(j)\big)j_1^2+E(j)
\end{align}
for some constant $d$ and some functional $E(j)$, and the prime here indicates a derivative with respect to $j$. Next, the coefficient of the monomial $j_1j_2^2j_4$ imposes $C'''(j)=0$, and so $C(j)$ is at most a quadratic functional of $j$. The coefficient of the monomial $j_1^2j_2j_4$ imposes the equation
\begin{align}
\label{eq:ell2to3polycases}
a(C''(j)-a^2/5)=0
\end{align}
and so either $a=0$, or $a\neq 0$ and $C''(j)=a^2/5$. We will treat these cases separately.

\sm

When $a=0$, $f_1(j,j_1) = \frac{1}{2}C(j)j_1^2+D(j)$, and $C(j)=C(0)+C'(0)j+\frac{1}{2}C''(0)j^2$. The coefficient of the monomial $j_1j_2j_4$ imposes the equation
\begin{align}
D^{(4)}(j)=\frac{3}{20}\frac{\partial^2}{\partial j^2}C(j)^2
\end{align}
which implies that $D(j)$ is at most a degree 6 polynomial functional of $j$. Then, the coefficient of the monomial $j_1j_4$ determines $E(j)$
\begin{align}
\label{eq:ell2to3polyh}
    E'''(j) = -\frac{21}{50}dC'(j)-\frac{21}{500}C(j)^2C'(j)+\frac{21}{50}C'(j)D''(j) + \frac{1}{2}AD'''(j) + \frac{7}{25}C(j)D'''(j)
\end{align}
The remaining constraints impose
\begin{align}
D'''(j)=\frac{3}{20}\frac{\partial}{\partial j}C(j)^2
\end{align}

\sm

And so when $a=0$, the most general polynomial order 2 Lagrangian which commutes with an order 3 Lagrangian is (up to a normalization)
\begin{align}
\ell_2(j,j_1,j_2)=&\,j_2^2 + \frac{1}{2}\big(C(0)+C'(0)j+\frac{1}{2}C''(0)j^2\big)j_1^2 \nonumber \\
&+ \frac{1}{800}j^2\big(5C(0)+C'(0)j\big)^2+\frac{1}{800}j^2\big(5C(0)+3C'(0)j+C''(0)j^2\big)^2 + q(j)
\end{align}
where $q(j)$ is an arbitrary quadratic functional of $j$. A quick check shows that these are precisely the order 2 Lagrangians which commute with the polynomial order 1 Lagrangians $\ell_1(j,j_1)$ in (\ref{eq:ell1poly}), as long as $C''(0)=\frac{5}{3}f^{(4)}(0)$ and $C'(0)=\frac{5}{3}f'''(0)$. The extra free parameter $C(0)$ appears in $\ell_2(j,j_1,j_2)$ in front of a term proportional to $\ell_1(j,j_1)$ and a term proportional to $j^2$, which trivially commute with $\ell_1(j,j_1)$, and so we may set $C(0)=0$ without loss of generality.

\sm

It is when $a\neq 0$ that we find the remaining polynomial order 2 Lagrangians which commute with an order 3 Lagrangian. We will find the order 2 and 3 $1d$ Lagrangians associated with the Bullough–Dodd model here. In this case, from equation (\ref{eq:ell2to3polycases}) we must have instead $C''(j)=a^2/5$. Next, the coefficient of the monomial $j_2j_4$ determines both $C(j)$ and $D(j)$
\begin{align}
    &C(j)=\frac{1}{10}(aj+b)^2, &&D(j)=\frac{1}{30000a^2}(aj+b)^6 + q(j)
\end{align}
with $b$ some new constant, and $q(j)$ some quadratic functional of $j$. Finally, the coefficient of the monomial $j_1j_4$ again determines $E(j)$ in terms of $C(j)$ and $D(j)$ through the equation (\ref{eq:ell2to3polyh}).

\sm

And so when $a\neq 0$, the most general polynomial order 2 Lagrangian which commutes with an order 3 Lagrangian is (up to a normalization)
\begin{align}
    \ell_2(j,j_1,j_2) = j_2^2 +\frac{1}{6}aj_1^3 + \frac{1}{20}(aj+b)^2j_1^2 + \frac{1}{30000a^2}(aj+b)^6 + q(j)
\end{align}
The order 2 $1d$ Lagrangian associated with the Bullough–Dodd model $\ell_{\text{BD}}=j_2^2-5j_1^3+45j^2j_1^2+27j^6$ may be obtained from this by setting $a=-30$ and $b=0$.

\subsubsection{The case $\frac{\partial^3\ell_2}{\partial j_2^3}\neq 0$}
If $\frac{\partial^3\ell_2}{\partial j_2^3}\neq 0$, then $\ell_2$ is not quadratic in $j_2$, and so cannot commute with an order 1 Lagrangian. The general solution to (\ref{eq:j8j4eq}) when $\frac{\partial^3\ell_2}{\partial j_2^3}\neq 0$ is
\begin{align}
\label{eq:2to3ell2}
    \ell_2(j,j_1,j_2) = \big(a j_2+b(j,j_1)\big)^{1/3}+c(j,j_1)
\end{align}
for some constant $a$, and functionals $b(j,j_1)$, and $c(j,j_1)$. Next, the coefficient of the monomial $j_8$ determines $f_2(j,j_1,j_2)$ in terms of $a$, $b(j,j_1)$ and $c(j,j_1)$
\begin{align}
\label{eq:f2ell2to3}
    f_2(j,j_1,j_2)=\frac{1}{2}\Big(\frac{\partial^2\ell_2}{\partial j_2^2}\Big)^{7/5}p_3(j,j_1,j_2) &+\frac{21\cdot 3^{1/5}(-1)^{2/5}}{10\cdot 2^{3/5}a^{6/5}}\big(aj_2+b(j,j_1)\big)^{4/3}c^{(0,2)}(j,j_1) \nonumber\\
    &+d\,\ell_2(j,j_1,j_2) + e(j,j_1)
\end{align}
where we have used the shorthand $c^{(n,m)}=\frac{\partial^n}{\partial j^n}\frac{\partial^m}{\partial j_1^m}c$, and $d$ is some constant and $e(j,j_1)$ is some functional. $p_3(j,j_1,j_2)$ is a degree three polynomial in $j_2$, whose coefficients depend on $a$ and $b(j,j_1)$ and their derivatives
\begin{align}
\label{eq:p3}
    p_3(j,j_1,j_2)&=\frac{21}{2a}j_2^3\,b^{(0,2)}-\frac{7}{10a^2}j_2^2\Big((b^{(0,1)})^2-24bb^{(0,2)}+3ab^{(1,0)}-21aj_1b^{(1,1)}\Big) \nonumber \\
    &+\frac{1}{10a^3}j_2\Big(21b(b^{(0,1)})^2 +36b^2b^{(0,2)}-57abb^{(1,0)}\nonumber \\&\hspace{3cm}-35aj_1b^{(0,1)}b^{(1,0)}+264aj_1bb^{(1,1)}+15a^2j_1^2b^{(2,0)}\Big) \nonumber \\
    &+\frac{1}{10a^4}\Big(18b^2(b^{(0,1)})^2-27b^3b^{(0,2)}-36ab^2b^{(1,0)}-15aj_1bb^{(0,1)}b^{(1,0)}\nonumber \\&\hspace{3cm}+117aj_1b^2b^{(1,1)}-10a^2j_1^2(b^{(1,0)})^2+15a^2j_1^2bb^{(2,0)}\Big)
\end{align}
Next, the coefficient of the monomial $j_3^3j_6$ implies
\begin{align}
    c^{(0,2)}(j,j_1)=0
\end{align}
This means that $c(j,j_1)$ is at most linear in $j_1$, but because it appears as an additive term in $\ell_2(j,j_1,j_2)$, it is therefore equivalent to a functional which only depends on $j$, i.e. we may write $c(j,j_1)=c(j)$. The last set of constraints comes from the coefficient of the monomial $j_3j_6$, the first of which being
\begin{align}
    &b^{(0,4)}(j,j_1)=0 && e^{(0,2)}(j,j_1)=0
\end{align}
and so $b(j,j_1)$ is a cubic polynomial in $j_1$, and we write
\begin{align}   
\label{eq:bjj1}
b(j,j_1)=a^{3/2}A(j)j_1^3+aB(j)j_1^2+a^{1/2}C(j)j_1+D(j)
\end{align}
the coefficient of $j_3j_6$ further imposes conditions on the functionals $A(j)$, $B(j)$, $C(j)$, and $D(j)$. These conditions take possibly the simplest form after making the redefinition
\begin{align}
\label{eq:Bredef}
    B(j)=\frac{C(j)^2- E(j)^2-2D'(j)}{3D(j)}
\end{align}
by defining some new functional $E(j)$. In terms of these functionals, the coefficient of the monomial $j_3j_6$ implies two coupled second order differential equations for $D(j)$ and $E(j)$
\begin{align}
\label{eq:DE1}
&30DD''-25D'^2+4E^4=0 \\
\label{eq:DE2}
&450D^2E''-150DD'E'+25D'^2E-4E^5=0
\end{align}
while $A(j)$ is fully determined in terms of $C(j)$, $D(j)$, and $E(j)$ via a quadratic equation
\begin{align}
\label{eq:A}
    \Big(A+\frac{3E^2C-C^3-3CD'+9C'D}{27D^2}\Big)^2-\frac{1}{3645\,D^4}(2E^3+15DE'-5D'E)^2=0
\end{align}
meanwhile, $C(j)$ is left undetermined. One can show that consistency between the coupled equations (\ref{eq:DE1}) and (\ref{eq:DE2}) requires that $D^{(7)}(j)=0$ and $E'''(j)=0$.

\sm

With no further constraints on $\ell_2$ and $\ell_3$, $c(j)=c(0)+c'(0)j+\frac{1}{2}c''(0)j^2$ and $e(j,j_1)=e(j)=e(0)+e'(0)j+\frac{1}{2}e''(0)j^2$ must be quadratic functionals of $j$. In this case, an 8 dimensional parameter space, with free parameters $a$, $c(0)$, $c'(0)$, $c''(0)$, $D(0)$, $D'(0)$, $E(0)$ and $E'(0)$, along with an undetermined functional $C(j)$, defines the solution space for $\ell_2(j,j_1,j_2)$.

\subsection{Definite scaling dimensions}
\label{ssec:scale}
The analysis in \autoref{ssec:ell12=0} and \autoref{ssec:ell23=0} is quite cumbersome, and is practically difficult to extend the analysis to higher order Lagrangians. Towards the aim of finding infinite dimensional mutually commuting subalgebras, which are relevant for integrable models, it is useful to further constrain the Lagrangians we consider. 

\sm

If we assign $x^-$ with a scaling dimension $[x^-]=-1$, and the current $j(x^-)$ with some scaling dimension $[j]$, which is not necessarily the canonical scaling dimension $1$, then a Lagrangian $\ell(\{j_i(x^-)\})$ with a definite scaling dimension $[\ell ]$ is an eigenfunction of the scaling operator $D_{[j]}$
\begin{align}
    &D_{[j]} \ell(\{j_i(x^-)\})= [\ell]\ell(\{j_i(x^-)\}) && \text{where}&&&D_{[j]}=\sum_{k=0}^{\infty}\big([j]+k\big)j_k(x^-)\frac{\partial}{\partial j_k(x^-)}
\end{align}
Restricting to Lagrangians with definite scaling dimension, the search for mutually commuting subalgebras becomes much easier. For scaling dimensions $[j]=2$ and $[j]=1$, we find the infinite sequences of mutually commuting Lagrangians associated with the well known integrable models of a single scalar, i.e. Korteweg–de Vries when $[j]=2$, and Liouville, sine-Gordon, modified Korteweg–de Vries, and Bullough–Dodd when $[j]=1$. For scaling dimensions $[j]=0,-1,-2$, we find new sequences of mutually commuting Lagrangians up to order 6, with no indication of the sequences being finite. Curiously, we were unable to find a sequence of mutually commuting Lagrangians beyond the scaling dimensions $[j]=0,\pm 1, \pm 2$.

\sm

In the following subsubsections, we simply list the sequences of mutually commuting Lagrangians up to order 6 for each scaling dimension mentioned above.
\subsubsection{$[j]=2$}
When $[j]=2$, we find an essentially unique sequence of mutually commuting Lagrangians (up to an affine transformation $j\to aj+b$ with $a,b\in\mathbb{C}$), which precisely correspond to the symmetries of the Korteweg–de Vries model, proven to be infinite dimensional originally in \cite{Kruskal2}
\begin{align}
\label{eq:KdVell1}
  &\ell_1:  &&j_1^2+2j^3,\\
  &\ell_2:  &&j_2^2+10jj_1^2+5j^4,\\
  &\ell_3:  &&j_3^2+14jj_2^2+70j^2j_1^2+14j^5,\\
  \label{eq:KdVell4}
  &\ell_4:  &&j_4^2+18jj_3^2-20j_2^3+126j^2j_2^2-35j_1^4+420j^3j_1^2+42j^6, \\
  &\ell_5:  &&j_5^2 +22jj_4^2-110j_2j_3^2+198j^2j_3^2-440jj_2^3-462j_1^2j_2^2\nonumber \\
    \label{eq:KdVell5}
  &  && +924j^3j_2^2-770jj_1^4 +2310j^4j_1^2+132j^7, \\
  \label{eq:KdVell6}
  &\ell_6: &&j_6^2 + 26 jj_5^2 - 182j_2j_4^2+286j^2j_4^2 - 2860jj_2j_3^2 - 858j_1^2j_3^2 \nonumber \\
  & && +1716j^3j_3^2 + 1001j_2^4 - 5720j^2j_2^3 - 12012jj_1^2j_2^2 + 6006j^4j_2^2 \nonumber \\
  & && - 10010j^2j_1^4 +12012j^5j_1^2 + 429j^8
\end{align}
The Korteweg–de Vries equation is generated by the action
\begin{align}
    S_{\text{KdV}} = \int d^2x\Big(\partial_+\phi\partial_-\phi - (\partial_-^2\phi)^2-2(\partial_-\phi)^3\Big)  
\end{align}
identifying the Korteweg–de Vries function $u=\partial_-\phi$, with the equation of motion
\begin{align}
    u_t-6uu_x+u_{xxx}=0
\end{align}
setting momentarily $x=x^-$, $t=x^+$, and the subscripts indicating derivatives. This action is literally of the form (\ref{eq:scalardeform}), deforming the free scalar action with $\ell_1$ in (\ref{eq:KdVell1}). This sequence has commuting order 1 and order 2 Lagrangians, and so belongs to the class of Lagrangians found in \autoref{ssec:ell12=0}, and forms a mutually commuting subalgebra, including Lagrangians $\ell_n$ for all $n\in \mathbb{N}$.

\sm

This theory has an additional invariance under the Galilean transformation 
\begin{align}
    &x'= x+6ct,&&t'= t,&&&u'(x,t)= u(x',t')+c
\end{align}
which is generated by the $1d$ Lagrangian $\ell_G=x^-j+3x^+j^2$ (restoring the original notation, and treating $x^+$ here as a parameter). This Lagrangian does not commute with the mutually commuting $1d$ Lagrangians $\ell_n$ in (\ref{eq:KdVell1})–(\ref{eq:KdVell6}) and so on, but has the important property that the commutator between $\ell_G$ and $\ell_n$ is proportional to $\ell_{n-1}$
\begin{align}
    \ell''\sim\frac{1}{2}\partial_-\mathcal{E}(\ell_G)\mathcal{E}(\ell_n)\sim\frac{1}{2}\mathcal{E}(\ell_n)\sim(2n+1)\ell_{n-1}
\end{align}
This observation is what was originally used to find an explicit formula for all Korteweg–de Vries charges in \cite{10.1063/1.1665232}. Here, this observation is reframed as a direct consequence of the structure of the Lie algebra itself. 
\subsubsection{$[j]=1$}
When $[j]=1$, we find two sequences of mutually commuting Lagrangians (again unique up to an affine transformation $j\to aj+b$ with $a,b\in\mathbb{C}$). The first sequence corresponds to the symmetries of Liouville, sine-Gordon, and modified Korteweg–de Vries, also proven to be infinite dimensional in \cite{Kruskal2}
\begin{align}
\label{eq:mKdVell1}
   &\ell_1: &&j_1^2+j^4, \\
   &\ell_2: &&j_2^2+10j^2j_1^2+2j^6,\\
   &\ell_3: &&j_3^2+14j^2j_2^2-7j_1^4+70j^4j_1^2+5j^8,\\
\label{eq:mKdVell4}
   &\ell_4: &&j_4^2+18j^2j_3^2-40jj_2^3-102j_1^2j_2^2+126j^4j_2^2-266j^2j_1^4+420j^6j_1^2+14j^{10}, \\
   &\ell_5: && j_5^2 + 22j^2j_4^2-220jj_2j_3^2-198j_1^2j_3^2+121j_2^4 + 198j^4j_3^2-880j^3j_2^3  \nonumber \\
\label{eq:mKdVell5}
   & &&- 4092j^2j_1^2j_2^2+\frac{1738}{5}j_1^6 +924j^6j_2^2-4466j^4j_1^4+2310j^8j_1^2+42j^{12},\\
   &\ell_6: && j_6^2 +26j^2j_5^2-364jj_2j_4^2-338j_1^2j_4^2 + 1846j_2^2j_3^2 +\frac{2548}{3}j_1j_3^3 + 286j^4j_4^2 \nonumber \\
   & && -5720j^3j_2j_3^2-8580j^2j_1^2j_3^2 + 7150j^2j_2^4 +19734j_1^4j_2^2 +36608j_1^2j_2^3 \nonumber \\
   & && +1716j^6j_3^2 - 77220j^4j_1^2j_2^2-11440j^5j_2^3 +31460j^2j_1^6 +6006j^8j_2^2 \nonumber \\
   & &&-52052j^6j_1^4 +12012j^{10}j_1^2 + 132j^{14}
\end{align}
In this case, along with actions of the form (\ref{eq:scalardeform}) which include the modified Korteweg–de Vries model, there are also Lorentz invariant actions invariant under these symmetries of the form 
\begin{align}
    S = \int d^2x\Big(\partial_+\phi\partial_-\phi+Ae^{2\phi}+Be^{-2\phi}\Big)
\end{align}
which include sinh-Gordon when $A=B$ and Liouville when $B=0$. The symmetries of the sine-Gordon model are obtained after the affine transformation $j\to \sqrt{-1}j$. A relationship exists between the sequence of Lagrangians associated with Korteweg–de Vries (\ref{eq:KdVell1})--(\ref{eq:KdVell5}) and modified Korteweg–de Vries (\ref{eq:mKdVell1})--(\ref{eq:mKdVell5}) \cite{Kruskal1}. Starting with $\ell_0=j^2$ and the Lagrangians (\ref{eq:KdVell1})--(\ref{eq:KdVell4}), performing the Miura transformation
\begin{align}
    j\to j_1 + j^2
\end{align}
we arrive at the Lagrangians (\ref{eq:mKdVell1})--(\ref{eq:mKdVell5}), up to total derivatives. This relationship proves that this sequence is infinite dimensional, and forms a mutually commuting subalgebra of Lagrangians $\ell_n$ for all $n\in \mathbb{N}$.

\sm

The second sequence of Lagrangians with $[j]=1$ corresponds to the symmetries of the Bullough–Dodd model, originally discovered in \cite{doi:10.1098/rspa.1977.0012} and proven to be infinite dimensional in \cite{Zhiber:1979am,Mikhailov1981}
\begin{align}
  &\ell_2:  &&j_2^2-5j_1^3+45j^2j_1^2+27j^6,\\
  &\ell_3:  &&j_3^2-21j_1j_2^2+63j^2j_2^2-21j_1^4-126j^2j_1^3+1134j^4j_1^2+243j^8, \\
  &\ell_5:  &&j_5^2-33j_1j_4^2+44j_3^3+99j^2j_4^2-990jj_2j_3^2-594j_1^2j_3^2+561j_2^4 +8118j_1^3j_2^2\nonumber \\
  &  && +3168jj_1j_2^3-1782j^2j_1j_3^2+\frac{24156}{5}j_1^6 - 58806j^2j_1^2j_2^2-15444j^3j_2^3 \nonumber \\
  &  &&+3564j^4j_3^2 + 44550j^2j_1^5-26730j^4j_1j_2^2 + 58806j^6j_2^2-267300j^4j_1^4 \nonumber \\
  &  && - 53460j^6j_1^3 + 481140j^8j_1^2 +26244j^{12},\\
  &\ell_6: &&j_6^2 -39j_1j_5^2+234j_3j_4^2 +117j^2j_5^2-1053j_1^2j_4^2 - 1638jj_2j_4^2 + 8424j_2^2j_3^2 \nonumber \\
  & && + 2691j_1j_3^3 -2808j^2j_1j_4^2 + 3393 j^2j_3^3 +24570jj_1j_2j_3^3 + 21177j_1^3j_3^2 \nonumber \\
  & && - 21762j_1j_2^4 + 5265j^4j_4^2 -123201j^2j_1^2j_3^2-103194j^3j_2j_3^2 + 121446j^2j_2^4 \nonumber \\
  & && +252369j_1^4j_2^2 + 531414j_1^2j_2^3 -66339 j^4j_1j_3^2 +324324j^3j_1j_2^3 +1400490j^2j_1^3j_2^2 \nonumber \\
  & && -85995j_1^7 + 116883j^6j_3^2 - 4624776j^4j_1^2j_2^2 -764478j^5j_2^3 + 
\frac{8612487}{5}j^2j_1^6 \nonumber \\
  & && -663390j^6j_1j_2^2 + 2321865j^4j_1^5 + 1393119j^8j_2^2 - 11609325j^6j_1^4 -995085j^8j_1^3\nonumber \\
  & && + 8955765 j^{10}j_1^2 + 295245j^{14}
\end{align}
In this case too, along with the actions of the form (\ref{eq:scalardeform}), there are Lorentz invariant actions invariant under these symmetries of the form
\begin{align}
    S=\int d^2x\Big(\partial_+\phi\partial_-\phi + Ae^{3\phi}+Be^{-6\phi}\Big)
\end{align}
which includes the Bullough–Dodd model when $A=2B$. This sequence forms an infinite dimensional mutually commuting subalgebra, including Lagrangians $\ell_n$ for all orders $n\not\equiv 1 \,\text{(mod 3)}$.
\subsubsection{$[j]=0$}
\label{sssec:j=0}
When $[j]=0$, we find a new, essentially unique sequence of mutually commuting Lagrangians (up to an affine transformation $j\to aj+b$ with $a,b\in\mathbb{C}$)
\begin{align}
\label{eq:j=0ell1}
    &\ell_1: &&\frac{j_1^2}{(1-j^2)^3}, \\
\label{eq:j=0ell2}
    &\ell_2: &&\frac{j_2^2}{(1-j^2)^5}-\frac{5}{3}\frac{1+8j^2}{(1-j^2)^7}j_1^4,\\
    &\ell_3: &&\frac{j_3^2}{(1-j^2)^7}-14\frac{j}{(1-j^2)^8}j_2^3-14\frac{2+13j^2}{(1-j^2)^9}j_1^2j_2^2 +\frac{14}{5}\frac{11+248j^2+416j^4}{(1-j^2)^{11}}j_1^6, \\
    &\ell_4: && \frac{j_4^2}{(1-j^2)^9} -75\frac{j}{(1-j^2)^{10}}j_2j_3^2-6\frac{10+59j^2}{(1-j^2)^{11}}j_1^2j_3^2 +\frac{37+448j^2}{(1-j^2)^{11}}j_2^4 \nonumber\\
    & && +12j\frac{334+1085j^2}{(1-j^2)^{12}}j_1^2j_2^3 +6\frac{311+6374j^2+10640j^4}{(1-j^2)^{13}}j_1^4j_2^2\nonumber \\
    & && -\frac{11}{7}\frac{967+45456j^2+213312j^4+170240j^6}{(1-j^2)^{15}}j_1^8, \\
    &\ell_5: && \frac{j_5^2}{(1-j^2)^{11}}-132\frac{j}{(1-j^2)^{12}}j_2j_4^2-22\frac{5+28j^2}{(1-j^2)^{13}}j_1^2j_4^2+\frac{22}{3}\frac{37+323j^2}{(1-j^2)^{13}}j_1j_3^3 \nonumber \\
    & && +22\frac{26+343j^2}{(1-j^2)^{13}}j_2^2j_3^2 + 132j\frac{253+826j^2}{(1-j^2)^{14}}j_1^2j_2j_3^2 - 154j\frac{42+179j^2}{(1-j^2)^{14}}j_2^5 \nonumber \\
    & && + 66\frac{89+1694j^2+2884j^4}{(1-j^2)^{15}}j_1^4j_3^2 - 22\frac{1187 + 34377j^2 + 68215j^4}{(1-j^2)^{15}}j_1^2j_2^4 \nonumber \\
    & && -572j\frac{2061+19138j^2+23156j^4}{(1-j^2)^{16}}j_1^4j_2^3 -572\frac{388 + 17259j^2+81228j^4+66500j^6}{(1-j^2)^{17}}j_1^6j_2^2 \nonumber \\
    & &&+\frac{1430}{9}\frac{941+76336j^2+701760j^4+1512448j^6+744800j^8}{(1-j^2)^{19}}j_1^{10},
    \end{align}
    \begin{align}
    \label{eq:j=0ell6}
    &\ell_6: &&\frac{j_6^2}{(1-j^2)^{13}} - 208\frac{j}{(1-j^2)^{14}}j_2j_5^2+182\frac{j}{(1-j^2)^{14}}j_4^3-26\frac{7+38j^2}{(1-j^2)^{15}}j_1^2j_5^2 \nonumber \\
    & &&+78\frac{23+197j^2}{(1-j^2)^{15}}j_1j_3j_4^2+26\frac{49+716j^2}{(1-j^2)^{15}}j_2^2j_4^2-\frac{13}{3}\frac{169+1700j^2}{(1-j^2)^{15}}j_3^4 \nonumber \\
    & &&+1352j\frac{59+196j^2}{(1-j^2)^{16}}j_1^2j_2j_4^2 - \frac{208}{3}j\frac{4121 + 16261j^2}{(1-j^2)^{16}}j_1j_2j_3^3 \nonumber \\
    & &&-52j\frac{4316+19803j^2}{(1-j^2)^{16}}j_2^3j_3^2 + 26\frac{599+10712j^2+18704j^4}{(1-j^2)^{17}}j_1^4j_4^2 \nonumber \\
    & &&-\frac{52}{3}\frac{5477+139337j^2+282842j^4}{(1-j^2)^{17}}j_1^3j_3^3 -52\frac{10333 + 309952j^2 + 633304j^4}{(1-j^2)^{17}}j_1^2j_2^2j_3^2 \nonumber\\
    & &&+ \frac{26}{5}\frac{7631 + 247380j^2+ 560266j^4}{(1-j^2)^{17}}j_2^6 -156j\frac{83319 + 778156j^2 + 973420j^4}{(1-j^2)^{18}}j_1^4j_2j_3^2 \nonumber \\
    & && +\frac{52}{5}j\frac{1710777 + 18463942j^2 +25032140j^4}{(1-j^2)^{18}}j_1^2j_2^5 \nonumber \\
    & &&-52\frac{17348 + 729759j^2 + 3482868j^4 + 2975560j^6}{(1-j^2)^{19}}j_1^6j_3^2 \nonumber \\
    & && + 26\frac{506621 + 28111390j^2 + 151882164j^4 + 140461440j^6}{(1-j^2)^{19}}j_1^4j_2^4 \nonumber \\
    & && + 104j\frac{4140512 + 76016221j^2 + 248907652j^4 + 167242880j^6}{(1-j^2)^{20}}j_1^6j_2^3 \nonumber \\
    & &&+26\frac{1}{(1-j^2)^{21}}\Big(1635523 + 129034676j^2 \nonumber \\ & && \hspace{3.5cm}+1193614224j^4 + 2627191952j^6 + 1336720000j^8\Big)j_1^8j_2^2 \nonumber \\
    & &&-\frac{442}{11}\frac{1}{(1-j^2)^{23}}\Big(617381 + 76885016j^2 + 1168922816j^4 \nonumber \\ 
    & &&\hspace{4cm}+ 4793764096j^6 + 6192510016j^8 + 2138752000j^{10}\Big)j_1^{12}
\end{align}

\sm
\begin{figure}
\centering
\begin{tikzpicture}
\begin{axis}
[
    xmin = -50, xmax = 50,
	ymin = -1.05, ymax = 1.05,
	xtick distance = 25,
	ytick distance = 0.5,
	grid = both,
	minor tick num = 0,
	major grid style = {lightgray!25},
	minor grid style = {lightgray!25},
	width = \textwidth,
	height = 0.5\textwidth,
	xlabel = {$z$},
 ]
]
\addplot [
    smooth,
    thick,
    color=blue,
    ] file[skip first] {j0waveplot.txt};
\addlegendentry{$u(z)$}
\end{axis}
\end{tikzpicture}
\caption{Wave profile of the single wave solution (\ref{eq:j=0wavesoln}) to the equation (\ref{eq:j=0KdV}).}
\end{figure}
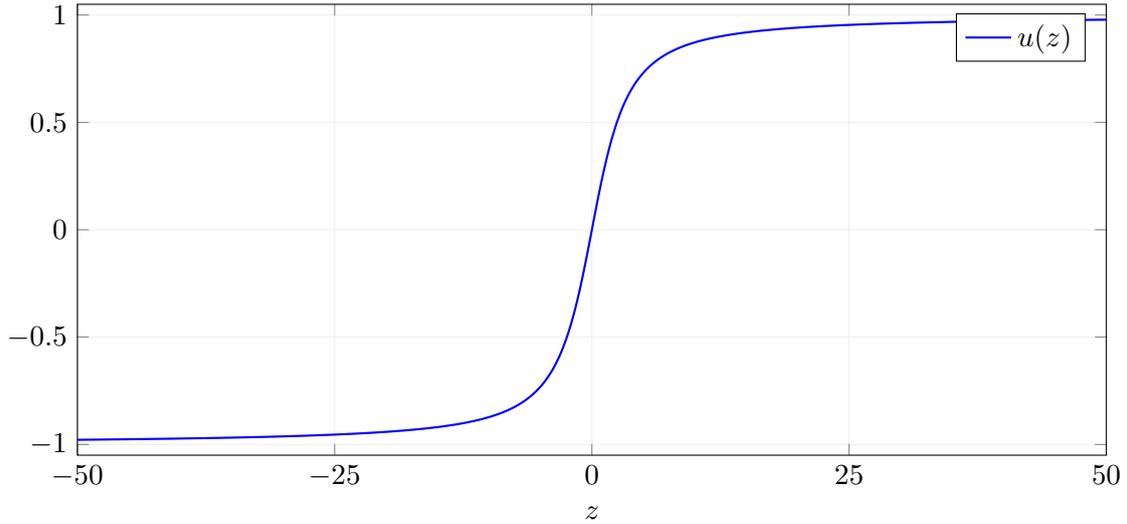
In this case, there is no previously known integrable model which has symmetries associated with these Lagrangians. The order 1 Lagrangian (\ref{eq:j=0ell1}) can be obtained from (\ref{eq:ell1abdform}). We have not proven that this sequence is infinite dimensional, but have found Lagrangians in this sequence up to order 6, with no indication of it terminating at finite order. Assuming this sequence is in fact infinite dimensional, it defines an infinite hierarchy of integrable models of the form (\ref{eq:scalardeform}), with the simplest action being
\begin{align}
    S=\int d^2x\Big(\partial_+\phi\partial_-\phi - \lambda\frac{(\partial_-^2\phi)^2}{(1-(\partial_-\phi)^2)^3}\Big)
\end{align}
with the equation of motion
\begin{align}
\label{eq:j=0KdV}
    u_t+3\lambda\frac{1+7u^2}{(1-u^2)^5}u_x^3+12\lambda\frac{u}{(1-u^2)^4}u_xu_{xx}+\lambda\frac{1}{(1-u^2)^3}u_{xxx}=0
\end{align}
where we have written it in the standard Korteweg–de Vries form with $u=\partial_-\phi$ and $x=x^-$, $t=x^+$. Wave solutions $u(x,t)=u(y)$ with $y=x+ct$ can be solved by quadrature. Integrating this equation with respect to $y$, we get
\begin{align}
\label{eq:j=0KdVwave}
    c\, u + 3\lambda \frac{u}{(1-u^2)^4}u_y^2+\lambda\frac{1}{(1-u^2)^3}u_{yy}=b
\end{align}
the order of this equation may be reduced, solving for $u_y$
\begin{align}
    u_y^2 = \frac{c}{\lambda}(u^2-1)^3(u-a)(u-b)
\end{align}
with new constants $a$ and $b$. One particularly symmetric wave solution is obtained by setting $a=1$ and $b=-1$, and is related to the hyperbolic Kepler equation, satisfying
\begin{align}
    &u(z) = \tanh{\frac{v(z)}{2}}, && v(z)+\sinh{v(z)} = z
\end{align}
where $z=\pm4\sqrt{\frac{c}{\lambda}}y+d$, with some constant $d$. This has an analytical solution \cite{Burniston_Siewert_1973}
\begin{align}
\label{eq:j=0wavesoln}
    u(z) = \frac{z/4}{1+\frac{z}{2\pi}\int_0^1 ds\arctan\Big(\frac{z}{\pi}+\frac{1}{\pi}\log\big(\frac{s}{1-s}\big)+\frac{1}{2\pi}\big(\frac{1}{s}-\frac{1}{1-s}\big)\Big)} 
\end{align}

\sm

Choosing different values of $a$ and $b$, one finds numerical solutions which ``look" like a nonlinear combination of waves (\ref{eq:j=0wavesoln}) at different separations moving with the same velocity. In contrast to the Korteweg–de Vries solitary waves, at far separation these solutions do not approach a naive linear superposition of waves (\ref{eq:j=0wavesoln}) of the form $u_{12}=u_1+u_2$, as they empirically maintain the condition $-1<u<1$. Multi-wave solutions can in principle be solved for in a manner described in detail for the original Korteweg–de Vries model in \cite{Laxorig}. In particular, it is possible to reduce the order of the differential equation for $N$-tuple wave solutions to one using the first $N$ conserved quantities defined by the sequence (\ref{eq:j=0ell1})–(\ref{eq:j=0ell6}), although the resulting equations are too complicated to solve here. 

\subsubsection{$[j]=-1$}
\label{sssec:j-1}
When $[j]=-1$, there is another new, essentially unique sequence of mutually commuting Lagrangians up to order 6 (up to an affine transformation $j_1\to aj_1+b$ with $a,b\in\mathbb{C}$)
\begin{align}
\label{eq:j=-1ell1}
    &\ell_1: &&(1+j_1^2)^{1/2}, \\
\label{eq:j=-1ell2}
    &\ell_2: &&\frac{j_2^2}{(1+j_1^2)^{5/2}},\\
    &\ell_3: &&\frac{j_3^2}{(1+j_1^2)^{7/2}}+\frac{7}{4}\frac{1-4j_1^2}{(1+j_1^2)^{11/2}}j_2^4,\\
    &\ell_4: &&\frac{j_4^2}{(1+j_1^2)^{9/2}}+10\frac{j_1}{(1+j_1^2)^{11/2}}j_3^3+\frac{3}{2}\frac{17-60j_1^2}{(1+j_1^2)^{13/2}}j_2^2j_3^2 +\frac{11}{40}\frac{79-1116j_1^2+1080j_1^4}{(1+j_1^2)^{17/2}}j_2^6, \\
    &\ell_5: &&\frac{j_5^2}{(1+j_1^2)^{11/2}} + 55\frac{j_1}{(1+j_1^2)^{13/2}}j_3j_4^2+\frac{33}{2}\frac{3-10j_1^2}{(1+j_1^2)^{15/2}}j_2^2j_4^2-\frac{11}{4}\frac{11-80j_1^2}{(1+j_1^2)^{15/2}}j_3^4 \nonumber \\
    & &&+286j_1\frac{8-15j_1^2}{(1+j_1^2)^{17/2}}j_2^2j_3^3 +\frac{429}{8}\frac{23-292j_1^2+280j_1^4}{(1+j_1^2)^{19/2}}j_2^4j_3^2 \nonumber \\
    & &&+\frac{143}{448}\frac{2339-73512j_1^2+219024j_1^4-100800j_1^6}{(1+j_1^2)^{23/2}}j_2^8, \\
    \label{eq:j=-1ell6}
    &\ell_6: &&\frac{j_6^2}{(1+j_1^2)^{13/2}} + 91\frac{j_1}{(1+j_1^2)^{15/2}}j_3j_5^2 + \frac{13}{2}\frac{13-42j_1^2}{(1+j_1^2)^{17/2}}j_2^2j_5^2 - \frac{637}{3}\frac{1-5j_1^2}{(1+j_1^2)^{17/2}}j_2j_4^3 \nonumber \\
    & &&-\frac{13}{2}\frac{71-560j_1^2}{(1+j_1^2)^{17/2}}j_3^2j_4^2 + \frac{13}{2}j_1\frac{2791-5250j_1^2}{(1+j_1^2)^{19/2}}j_2^2j_3j_4^2 -\frac{65}{2}j_1\frac{111-280j_1^2}{(1+j_1^2)^{19/2}}j_3^5 \nonumber \\
    & && +\frac{39}{8}\frac{733-8632j_1^2+8400j_1^4}{(1+j_1^2)^{21/2}}j_2^4j_4^2 - \frac{13}{8}\frac{9517-176160j_1^2+201600j_1^4}{(1+j_1^2)^{21/2}}j_2^2j_3^4 \nonumber \\
    & &&+ \frac{221}{4}j_1\frac{9021-52600j_1^2+36400j_1^4}{(1+j_1^2)^{23/2}}j_2^4j_3^3 \nonumber \\
    & &&+\frac{221}{16}\frac{7547-220008j_1^2+647760j_1^4-302400j_1^6}{(1+j_1^2)^{25/2}}j_2^6j_3^2 \nonumber \\
    & && +\frac{29393}{384}\frac{677-37816j_1^2+230112j_1^4-309120j_1^6 + 86400j_1^8}{(1+j_1^2)^{29/2}}j_2^{10}
\end{align}

\sm

In this case too, there is no previously known integrable model which has symmetries associated with these Lagrangians. The order 1 Lagrangian (\ref{eq:j=-1ell1}) can be obtained from (\ref{eq:ell1abdform}). Assuming that this sequence is infinite dimensional, it defines another infinite hierarchy of integrable models, with the simplest action being
\begin{align}
    S=\int d^2x\Big(\partial_+\phi\partial_-\phi-\lambda \big(1+(\partial_-^2\phi)^2\big)^{1/2}\Big)
\end{align}
with the equation of motion
\begin{align}
\label{eq:j=-1KdV}
    u_t -\frac{3\lambda}{2}\frac{u_xu_{xx}^2}{(1+u_x^2)^{5/2}} + \frac{\lambda}{2}\frac{u_{xxx}}{(1+u_x^2)^{3/2}}=0
\end{align}
written in the standard Korteweg–de Vries form. Wave solutions $u(x,t)=u(y)$ with $y=x+ct$ can be solved by quadrature. Integrating this equation with respect to $y$, we get
\begin{align}
\label{eq:j=-1KdVwave}
    c\,u+\frac{\lambda}{2}\frac{u_{yy}}{(1+u_y^2)^{3/2}}=b
\end{align}
the order of this equation may be reduced, solving for $u_y$
\begin{align}
    u_y^2 = \frac{1}{c^2} \frac{\lambda^2-c^2((u-a)(u-b))^2}{((u-a)(u-b))^2}
\end{align}
with new constants $a$ and $b$. Perhaps the simplest wave solution is obtained when $a,b=0$, which after integrating satisfies the equation
\begin{align}
\label{eq:j=-1wavesoln}
  &u(z)= \sqrt{\frac{\lambda}{c}}\sin v(z)&& E(v(z),-1)-F(v(z),-1) = z
\end{align}
where $z=\pm\sqrt{\frac{c}{\lambda}}y+d$, with some constant $d$. Here $F(\varphi,m)=\int_0^{\varphi}d\theta(1-m\sin^2\theta)^{-1/2}$ and $E(\varphi,m)=\int_0^{\varphi}d\theta(1-m\sin^2\theta)^{1/2}$ are the incomplete elliptic integrals of the first and second kind, respectively. $u(z)$ is periodic with a period of $\frac{4\sqrt{\pi}\Gamma(7/4)}{3\Gamma(5/4)}$.

\sm

As before, multi-wave solutions can in principle be solved for in a manner described in \cite{Laxorig}. Here, it is possible to reduce the order of the differential equation for $N$-tuple wave solutions to one using the first $N$ conserved quantities defined by the sequence (\ref{eq:j=-1ell1})–(\ref{eq:j=-1ell6}), although the resulting equations are too complicated to solve here.

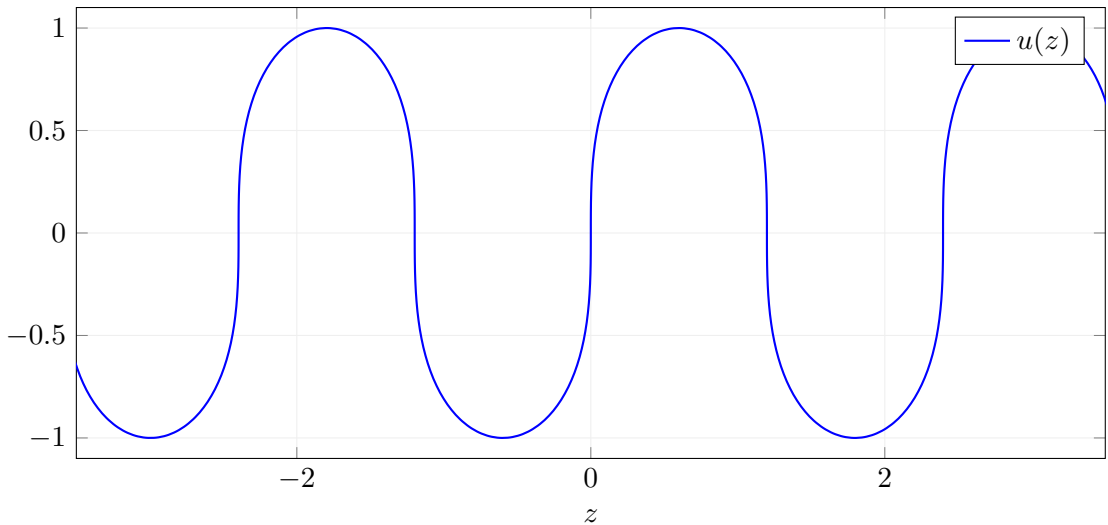
\begin{figure}
\centering
\begin{tikzpicture}
\begin{axis}
[
    xmin = -3.5, xmax = 3.5,
	ymin = -1.1, ymax = 1.1,
	xtick distance = 2,
	ytick distance = 0.5,
	grid = both,
	minor tick num = 0,
	major grid style = {lightgray!25},
	minor grid style = {lightgray!25},
	width = \textwidth,
	height = 0.5\textwidth,
	xlabel = {$z$},
 ]
]
\addplot [
    smooth,
    thick,
    color=blue,
    ] file[skip first] {j-1waveplot.txt};
\addlegendentry{$u(z)$}
\end{axis}
\end{tikzpicture}
\caption{Wave profile of the single wave solution (\ref{eq:j=-1wavesoln}) to the equation (\ref{eq:j=-1KdV}) with $\sqrt{c/\lambda}=1$.}
\end{figure}
\subsubsection{$[j]=-2$}
\label{sssec:j-2}
When $[j]=-2$, there is another new, essentially unique sequence of mutually commuting Lagrangians up to order 6 (up to an affine transformation $j_2\to aj_2+b$ with $a,b\in\mathbb{C}$) 
\begin{align}
\label{eq:j=-2ell2}
    &\ell_2: &&(1+j_2)^{1/3}, \\
\label{eq:j=-2ell3}
    &\ell_3: &&\frac{j_3^2}{(1+j_2)^{7/3}},\\
    &\ell_5: &&\frac{j_5^2}{(1+j_2)^{11/3}} + \frac{44}{9}\frac{j_4^3}{(1+j_2)^{14/3}}-\frac{220}{9}\frac{j_3^2j_4^2}{(1+j_2)^{17/3}} + \frac{6545}{243}\frac{j_3^6}{(1+j_2)^{23/3}}, \\
    \label{eq:j=-2ell6}
    &\ell_6: &&\frac{j_6^2}{(1+j_2)^{13/3}}+26\frac{j_4j_5^2}{(1+j_2)^{16/3}}-\frac{130}{3}\frac{j_3^2j_5^2}{(1+j_2)^{19/3}}+\frac{1456}{27}\frac{j_4^4}{(1+j_2)^{19/3}} \nonumber \\
    & &&-\frac{49400}{81}\frac{j_3^2j_4^3}{(1+j_2)^{22/3}}+\frac{108680}{81}\frac{j_3^4j_4^2}{(1+j_2)^{25/3}}-\frac{2377375}{2187}\frac{j_3^8}{(1+j_2)^{31/3}}
\end{align}

\sm

In this case too, there is no previously known integrable model which has symmetries associated with these Lagrangians. The order 2 Lagrangian (\ref{eq:j=-2ell2}) can be obtained from (\ref{eq:2to3ell2}). This sequence appears to follow the same pattern as the sequence corresponding the Bullough–Dodd model, missing Lagrangians $\ell_n$ with order $n \equiv 1\,\text{(mod 3)}$. Assuming that this sequence is infinite dimensional, it defines another infinite hierarchy of integrable models, with the simplest action being
\begin{align}
    S=\int d^2x\Big(\partial_+\phi\partial_-\phi-\lambda \big(1+\partial_-^3\phi\big)^{1/3}\Big)
\end{align}
with the equation of motion
\begin{align}
    u_t+\frac{40\lambda}{81}\frac{u_{xxx}^3}{(1+u_{xx})^{11/3}}-\frac{5\lambda}{9}\frac{u_{xxx}u_{xxxx}}{(1+u_{xx})^{8/3}}+\frac{\lambda}{9}\frac{u_{xxxxx}}{(1+u_{xx})^{5/3}}=0
\end{align}
written in the standard Korteweg–de Vries form. This partial differential can be reduced to a first order ordinary differential equation after considering wave solutions $u(x,t)=u(y)$ with $y=x+ct$. In terms of $y$, this equation becomes
\begin{align}
\label{eq:j=-2waveeqn}
     cu_y+\frac{40\lambda}{81}\frac{u_{yyy}^3}{(1+u_{yy})^{11/3}}-\frac{5\lambda}{9}\frac{u_{yyy}u_{yyyy}}{(1+u_{yy})^{8/3}}+\frac{\lambda}{9}\frac{u_{yyyyy}}{(1+u_{yy})^{5/3}}=0
\end{align}
To reduce this fifth order ordinary differential equation to first order, one notices that because of the sequence of symmetries (\ref{eq:j=-2ell2})–(\ref{eq:j=-2ell6}), there are several integrating factors of the equation (\ref{eq:j=-2waveeqn}). The integrating factors which are sufficient to reduce this equation to first order are $1$, $u$, and
\begin{align}
    \frac{u_{yyyyyy}}{(1+u_{yy})^{7/3}}-7\frac{u_{yyy}u_{yyyyy}}{(1+u_{yy})^{10/3}}-\frac{14}{3}\frac{u_{yyyy}^4}{(1+u_{yy})^{10/3}}+\frac{245}{9}\frac{u_{yyy}^2u_{yyyy}}{(1+u_{yy})^{13/3}}-\frac{455}{27}\frac{u_{yyy}^4}{(1+u_{yy})^{16/3}}
\end{align}
Multiplying (\ref{eq:j=-2waveeqn}) separately by each of these integrating factors, it is possible to reduce the equation to an autonomous second order equation, which can be reduced to a first order equation in the standard way. The resulting equation is complicated and can in principle be solved, but we will not continue this analysis here.
\section{Quantization}
\label{sec:quantum}
In this section, we consider how the classical Lie algebras, and their mutually commuting subalgebras, discussed in this paper are modified after quantization. Nonlinear symmetries are deformed under quantization, and so we expect that these Lie algebras, which generate highly nonlinear infinitesimal symmetry transformations, will be deformed dramatically. In the context of integrable models, an important question is whether the mutually commuting subalgebras of the classical Lie algebra still exist in some form after quantization. For simplicity, we will restrict to considering the Lie algebra associated with a single scalar as in \autoref{sec:deform}.

\sm

For the quantum theory it is more convenient to work directly in Euclidean space with complex coordinates $z$ and $\bar{z}$. To keep track of the quantum corrections, we introduce the parameter $\alpha'$, normalizing the single scalar action as follows
\begin{align}
    S=\frac{1}{2\pi\alpha'}\int d^2z\bar{\partial}\phi\partial\phi
\end{align}
We take the generators of the Lie algebra to be charges $Q_\ell$ of the form
\begin{align}
\label{eq:quantumQ}
    Q_\ell=-\frac{1}{\alpha'}\oint_C\frac{dz}{2\pi i}\colon\hspace{-.1cm}\ell(\{j_i(z)\},z)\colon\hspace{-.1cm}
\end{align}
where $\colon\hspace{-.1cm}\ell(\{j_i(z)\},z)\colon\hspace{-.1cm}$ is a normal ordered $1d$ Lagrangian operator, that is a functional of the holomorphic current $j(z)=\partial\phi$. Normal ordering is defined in the standard way, such that for instance
\begin{align}
    \colon \hspace{-.1cm}\phi(z,\bar{z})\colon \hspace{-.1cm}&= \phi(z,\bar{z}) \\
    \colon \hspace{-.1cm}\phi(z_1,\bar{z}_1)\phi(z_2,\bar{z}_2)\colon \hspace{-.1cm}&= \phi(z_1,\bar{z}_1)\phi(z_2,\bar{z}_2)-\langle \phi(z_1,\bar{z}_1)\phi(z_2,\bar{z}_2)\rangle
\end{align}
where $\langle \phi(z_1,\bar{z}_1)\phi(z_2,\bar{z}_2)\rangle=-\frac{\alpha'}{2}\log|z_{12}|^2$ with $z_{12}=z_1-z_2$. Normal ordering is defined similarly for an arbitrary product of fields. With normal ordering defined this way, $\colon\hspace{-.1cm}\ell(\{j_i(z)\},z)\colon\hspace{-.1cm}$ is holomorphic. The product of two normal ordered $1d$ Lagrangian operators $\colon\hspace{-.1cm}\ell(\{j_i(z_1)\},z_1)\colon\hspace{-.1cm}$ and $\colon\hspace{-.1cm}\ell'(\{j_{i'}(z_2)\},z_2)\colon\hspace{-.1cm}$ may be written in normal ordered form
\begin{align}
    \colon\hspace{-.1cm}\ell(z_1)\colon\hspace{-.1cm}\colon\hspace{-.1cm}\ell'(z_2)\colon\hspace{-.1cm} = \colon\hspace{-.1cm}\sum_{N=0}^{\infty}\frac{1}{N!}\sum_{\substack{n_i,m_i=0 \\ i=1,\dots,N}}^{\infty}\frac{\partial^N\ell(z_1)}{\partial j_{n_1}\cdots\partial j_{n_N}}\prod_{i=1}^N\langle j_{n_i}(z_1)j_{m_i}(z_2)\rangle\frac{\partial^N\ell'(z_2)}{\partial j_{m_1}\cdots\partial j_{m_N}}\colon\hspace{-.1cm}
\end{align}
In the quantum theory, it is again more convenient to describe the Lie algebra in terms of the commutator $[Q_\ell,Q_{\ell'}]$, which we may readily compute using the above expression
\begin{align}
     &\hspace{-2.3cm}[Q_\ell,Q_{\ell'}]=Q_{\ell''},\\
    \label{eq:quantumscalarLie}
    \ell''=-\frac{1}{\alpha'}\sum_{N=1}^{\infty}&\frac{1}{N!}\Big(-\frac{\alpha'}{2}\Big)^N\sum_{\substack{n_i,m_i=0 \\ i=1,\dots,N}}^{\infty}\frac{\prod_{i=1}^N(n_i+m_i+1)!}{\big(2N-1+\sum_{i=1}^N(n_i+m_i)\big)!}\times \nonumber \\
    &\times\Big((-1)^{\sum_{i=1}^N(n_i+m_i)}\partial^{2N-1+\sum_{i=1}^Nn_i}\frac{\partial^N\ell}{\partial j_{n_1}\cdots\partial j_{n_N}}\partial^{\sum_{i=1}^Nm_i}\frac{\partial^N\ell'}{\partial j_{m_1}\cdots\partial j_{m_N}}\Big)
\end{align}
generating an infinite series of corrections to the commutator Lagrangian $\ell''$ in the parameter $\alpha'$. A quick inspection of (\ref{eq:quantumscalarLie}) reveals that the $N=1$ term is the classical contribution $\frac{1}{2}\partial\mathcal{E}(\ell)\mathcal{E}(\ell')$, while the $N>1$ terms are quantum corrections. For polynomial $1d$ Lagrangians $\ell$ and $\ell'$, the sum in (\ref{eq:quantumscalarLie}) terminates at a finite $N$, but in general one must consider the quantum corrections to all orders to describe the full Lie algebra.

\sm

Among other subalgebras, the Witt algebra (\ref{eq:scalarWitt}) is deformed, becoming the Virasoro algebra with central charge $c=1$, with Lagrangian generators $l_n = z^{n+1}T(z)$, as expected. The commutator Lagrangian $\ell''$ between $\ell=l_n$ and $\ell'=l_m$ is
\begin{align}
    \ell'' = (n-m)l_{n+m} - \frac{\alpha'}{12}(n^3-n)z^{n+m-1}
\end{align}
which upon contour integration (\ref{eq:quantumQ}) gives the Virasoro algebra in terms of the canonical Virasoro generators $L_n=Q_{l_n}$. 

\sm

What is the fate of the mutually commuting subalgebras of the classical Lie algebra after quantization? It is known that for the well known integrable models involving a single scalar, which have symmetries generated by polynomial $1d$ Lagrangians, that the mutually commuting Lagrangians are in fact modified, but only mildly, 
the coefficients of each monomial being renormalized, and they continue to commute quantum mechanically \cite{Kulish:1976ef,Sasaki:1987mm,Bazhanov:1994ft}. We will not explore polynomial $1d$ Lagrangians further in this section.

\sm

The situation is drastically different for non-polynomial $1d$ Lagrangians. Unless $\ell$ or $\ell'$ is a polynomial $1d$ Lagrangian, $Q_{\ell''}=0$ is no longer a polynomial equation in $j_k(z)$ where $k>i,i'$, and in general involves derivatives of $j(z)$ at all orders. Consequently, for two non-polynomial finite order $1d$ Lagrangians $\ell$ and $\ell'$ to commute quantum mechanically, they must satisfy effectively infinitely many equations at finite $\alpha'$. It is untenable therefore to expect classically commuting non-polynomial $1d$ Lagrangians to maintain their form after quantization. In fact we find that quantum corrections generically increase the order of the classically commuting non-polynomial $1d$ Lagrangians.

\sm

Solving $Q_{\ell''}=0$ order by order in $\alpha'$, starting with classically commuting $1d$ Lagrangians $\ell,\ell' = \ell_{\text{classical}}+\mathcal{O}(\alpha')$, we may solve for each $\alpha'^n$ correction to $\ell$ and $\ell'$ in a straightforward manner. In each of the three cases of classically commuting non-polynomial $1d$ Lagrangians with definite scaling dimension we found in \autoref{sssec:j=0}, \autoref{sssec:j-1}, and \autoref{sssec:j-2}, we compute the first order $\alpha'$ corrections to the first two Lagrangians in each sequence. Surprisingly, in each case we find, apart from the possibility of adding to each Lagrangian  at first order in $\alpha'$ a term proportional to one of the other Lagrangians in the sequence, a three parameter space of possible corrections at first order in $\alpha'$. This is in stark contrast from what was found classically, where there was an essentially unique sequence of commuting Lagrangians.

\sm

For the sequence we found with $[j]=0$, the first two Lagrangians $\ell_1$ (\ref{eq:j=0ell1}) and $\ell_2$ (\ref{eq:j=0ell2}) become order 2 and 3 Lagrangians, respectively, at first order in $\alpha'$. They are equal to
\begin{align}
    & \ell_1: &&\frac{j_1^2}{(1-j^2)^3}+\alpha'\Big(\frac{21}{10}\frac{j^4}{(1-j^2)^6}j_2^2-\frac{1}{30}\frac{1407-1212j^2+638j^4+574j^6}{(1-j^2)^8}j_1^4\Big) \nonumber\\
    & &&+c_1\alpha'\Big(\frac{j}{(1-j^2)^6}j_2^2-\frac{1}{3}j\frac{19+48j^2}{(1-j^2)^8}j_1^4\Big) +c_2\alpha'\Big(\frac{j^2}{(1-j^2)^6}j_2^2-\frac{1}{3}\frac{67-40j^2+40j^4}{(1-j^2)^8}j_1^4\Big) \nonumber\\
    & &&+c_3\alpha'\frac{j_1^4}{(1-j^2)^7} +\mathcal{O}(\alpha'^2),\\
    & \ell_2: &&\frac{j_2^2}{(1-j^2)^5}-\frac{5}{3}\frac{1+8j^2}{(1-j^2)^7}j_1^4 \nonumber\\
    & &&+\alpha'\Big(\frac{1}{28}j^2\frac{15+83j^2}{(1-j^2)^8}j_3^2-\frac{1}{252}j\frac{237+4454j^2+9421j^4}{(1-j^2)^9}j_2^3 \nonumber\\
    & &&\hspace{1cm}-\frac{1}{42}\frac{19668-14043j^2+16423j^4+18524j^6}{(1-j^2)^{10}}j_1^2j_2^2 \nonumber\\
    & && \hspace{1cm}+\frac{1}{90}\frac{51403+607397j^2-312564j^4+418844j^6+206380j^8}{(1-j^2)^{12}}j_1^6\Big) \nonumber\\
    & &&+c_1\alpha'\Big(\frac{5}{3}\frac{j}{(1-j^2)^8}j_3^2-\frac{5}{3}\frac{1+15j^2}{(1-j^2)^9}j_2^3-\frac{20}{3}j\frac{17+52j^2}{(1-j^2)^{10}}j_1^2j_2^2 \nonumber\\
    & &&\hspace{1.5cm}+4j\frac{69+592j^2+624j^4}{(1-j^2)^{12}}j_1^6\Big) \nonumber\\
    & &&+c_2\alpha'\Big(\frac{5}{3}\frac{j^2}{(1-j^2)^8}j_3^2 - \frac{10}{3}j\frac{1+7j^2}{(1-j^2)^9}j_2^3 - \frac{5}{3}\frac{135-41j^2+182j^4}{(1-j^2)^{10}}j_1^2j_2^2 \nonumber \\
    & &&\hspace{1.5cm}+\frac{1}{3}\frac{833+10467j^2-1704j^4+5824j^6}{(1-j^2)^{12}}j_1^6\Big) \nonumber\\
    & &&+c_3\alpha'\Big(10\frac{1}{(1-j^2)^9}j_1^2j_2^2-12\frac{1+12j^2}{(1-j^2)^{11}}j_1^6\Big) +\mathcal{O}(\alpha'^2)
\end{align}
For the sequence we found with $[j]=-1$, the first two Lagrangians $\ell_1$ (\ref{eq:j=-1ell1}) and $\ell_2$ (\ref{eq:j=-1ell2}) become order 3 and 4 Lagrangians, respectively, at first order in $\alpha'$. They are equal to
\begin{align}
    & \ell_1: &&(1+j_1^2)^{1/2}+\alpha'\Big(-\frac{1}{80}\frac{j_1^2}{(1+j_1^2)^{7/2}}j_3^2-\frac{7}{480}j_1^2\frac{6-5j_1^2}{(1+j_1^2)^{11/2}}j_2^4\Big)\nonumber\\
    & &&+c_1\alpha'\Big(\frac{1}{(1+j_1^2)^4}j_3^2+\frac{9}{4}\frac{1-4j_1^2}{(1+j_1^2)^6}j_2^4\Big)+c_2\alpha'\Big(\frac{j_1}{(1+j_1^2)^4}j_3^2+\frac{1}{4}j_1\frac{17-28j_1^2}{(1+j_1^2)^6}j_2^4\Big)\nonumber\\
    & &&+c_3\alpha'\frac{1}{(1+j_1^2)^{11/2}}j_2^4 +\mathcal{O}(\alpha'^2),
\end{align}
\begin{align}
    & \ell_2: &&\frac{j_2^2}{(1+j_1^2)^{5/2}}\nonumber\\
    & &&+\alpha'\Big(-\frac{11}{336}\frac{j_1^2}{(1+j_1^2)^{9/2}}j_4^2+\frac{1}{1008}j_1\frac{57-224j_1^2}{(1+j_1^2)^{11/2}}j_3^3 + \frac{1}{336}\frac{32-647j_1^2+812j_1^4}{(1+j_1^2)^{13/2}}j_2^2j_3^2 \nonumber\\
    & &&\hspace{1cm}+\frac{1}{10080}\frac{2866-53460j_1^2+153765j_1^4-68040j_1^6}{(1+j_1^2)^{17/2}}j_2^6\Big) \nonumber \\
    & &&+c_1\alpha'\Big(\frac{10}{3}\frac{1}{(1+j_1^2)^5}j_4^2 + \frac{110}{3}\frac{j_1}{(1+j_1^2)^6}j_3^3 +\frac{10}{3}\frac{29-109j_1^2}{(1+j_1^2)^7}j_2^2j_3^2 \nonumber\\
    & &&\hspace{1.5cm}+ \frac{7}{6}\frac{77-1152j_1^2+1212j_1^4}{(1+j_1^2)^9}j_2^6\Big) \nonumber\\
    & && + c_2\alpha'\Big(\frac{10}{3}\frac{j_1}{(1+j_1^2)^5}j_4^2-\frac{10}{3}\frac{1-10j_1^2}{(1+j_1^2)^6}j_3^3 + 20j_1\frac{8-15j_1^2} {(1+j_1^2)^7}j_2^2j_3^2 \nonumber\\
    & &&\hspace{1.5cm}+\frac{1}{6}j_1\frac{1691-9456j_1^2+5940j_1^4}{(1+j_1^2)^9}j_2^6\Big) \nonumber\\
    & &&+c_3\alpha'\Big(20\frac{1}{(1+j_1^2)^{13/2}}j_2^2j_3^2+2\frac{11-78j_1^2}{(1+j_1^2)^{17/2}}j_2^6\Big) + \mathcal{O}(\alpha'^2)
\end{align}
And finally, for the sequence we found with $[j]=-2$, the first two Lagrangians $\ell_2$ (\ref{eq:j=-2ell2}) and $\ell_3$ (\ref{eq:j=-2ell3}) become at most order 5 and 6 Lagrangians, respectively, at first order in $\alpha'$. They are equal to
\begin{align}
    & \ell_2: &&(1+j_2)^{1/3} +\alpha'\Big(\frac{25}{1701}\frac{j_3^2j_4^2}{(1+j_2)^{17/3}}-\frac{1585}{61236}\frac{j_3^6}{(1+j_2)^{23/3}}\Big) \nonumber\\
    & &&+c_1\alpha'\Big(\frac{j_5^2}{(1+j_2)^4}+\frac{47}{9}\frac{j_4^3}{(1+j_2)^5}-\frac{250}{9}\frac{j_3^2j_4^2}{(1+j_2)^6}+\frac{8204}{243}\frac{j_3^6}{(1+j_2)^8}\Big) \nonumber\\
    & && + c_2\alpha'\Big(\frac{j_5^2}{(1+j_2)^3}+\frac{38}{9}\frac{j_4^3}{(1+j_2)^4}-\frac{181}{9}\frac{j_3^2j_4^2}{(1+j_2)^5} + \frac{22894}{1215}\frac{j_3^6}{(1+j_2)^7}\Big) \nonumber\\
    & &&+ c_3\alpha'\Big(\frac{j_4^3}{(1+j_2)^{14/3}}-2\frac{j_3^2j_4^2}{(1+j_2)^{17/3}}+\frac{164}{135}\frac{j_3^6}{(1+j_2)^{23/3}}\Big) +\mathcal{O}(\alpha'^2), 
    \end{align}
    \begin{align}
    & \ell_3: &&\frac{j_3^2}{(1+j_2)^{7/3}}\nonumber\\
    & &&+\alpha'\Big(\frac{5}{1944}\frac{j_4j_5^2}{(1+j_2)^{16/3}}-\frac{2195}{11664}\frac{j_3^2j_5^2}{(1+j_2)^{19/3}} +\frac{455}{4374}\frac{j_4^4}{(1+j_2)^{19/3}} \nonumber\\
    & &&\hspace{1cm}-\frac{79415}{26244}\frac{j_3^2j_4^3}{(1+j_2)^{22/3}}+\frac{268265}{26244}\frac{j_3^4j_4^2}{(1+j_2)^{25/3}}-\frac{413329025}{39680928}\frac{j_3^8}{(1+j_2)^{31/3}}\Big) \nonumber\\
    & &&+c_1\alpha'\Big(-\frac{63}{5}\frac{j_6^2}{(1+j_2)^{14/3}}-\frac{1722}{5}\frac{j_4j_5^2}{(1+j_2)^{17/3}}+\frac{3017}{5}\frac{j_3^2j_5^2}{(1+j_2)^{20/3}} \nonumber\\
    & &&\hspace{1.5cm}-\frac{11228}{15}\frac{j_4^4}{(1+j_2)^{20/3}}+\frac{79625}{9}\frac{j_3^2j_4^3}{(1+j_2)^{23/3}}-\frac{181426}{9}\frac{j_3^4j_4^2}{(1+j_2)^{26/3}}\nonumber\\
    & && \hspace{1.5cm}+\frac{1413581}{81}\frac{j_3^8}{(1+j_2)^{32/3}}\Big)\nonumber\\
    & &&+c_2\alpha'\Big(-\frac{63}{5}\frac{j_6^2}{(1+j_2)^{11/3}}-294\frac{j_4j_5^2}{(1+j_2
    )^{14/3}} + 469\frac{j_3^2j_5^2}{(1+j_2)^{17/3}}-\frac{8456}{15}\frac{j_4^4}{(1+j_2)^{17/3}} \nonumber\\
    & &&\hspace{1.5cm}+\frac{273721}{45}\frac{j_3^2j_4^3}{(1+j_2)^{20/3}}-\frac{116221}{9}\frac{j_3^4j_4^2}{(1+j_2)^{23/3}}+\frac{757211}{81}\frac{j_3^8}{(1+j_2)^{29/3}}\Big) \nonumber\\
    & &&+c_3\alpha'\Big(-\frac{189}{5}\frac{j_4j_5^2}{(1+j_2)^{16/3}} + \frac{126}{5}\frac{j_3^2j_5^2}{(1+j_2)^{19/3}}-\frac{504}{5}\frac{j_4^4}{(1+j_2)^{19/3}} + \frac{3297}{5}\frac{j_3^2j_4^3}{(1+j_2)^{22/3}}\nonumber\\
    & &&\hspace{1.5cm} -\frac{4662}{5}\frac{j_3^4j_4^2}{(1+j_2)^{25/3}} + \frac{4838}{9}\frac{j_3^8}{(1+j_2)^{31/3}}\Big) +\mathcal{O}(\alpha'^2)
\end{align}

\sm

Importantly, it is still possible to find commuting non-polynomial $1d$ Lagrangians, at least to first order in $\alpha'$. If we assign $\alpha'$ with a dimension $[\alpha']=2[j]-2$, the Lagrangians continue to have a definite scaling dimension. Furthermore if $\ell$ and $\ell'$ have definite scaling dimensions, then $\ell''$ has a definite scaling dimension as long as $[\alpha']=2[j]-2$. Dimensional analysis then implies that when $[j]\leq 0$, the order of a given Lagrangian $\ell$ with definite scaling dimension will increase at most by $1-[j]$ at every order in $\alpha'$. Consequently, we expect in general that at finite $\alpha'$ the commuting non-polynomial $1d$ Lagrangians above will be essentially non-local. In the context of integrable models, the existence of even one quantum non-local charge is often sufficient to demonstrating integrability \cite{Luscher:1977uq,Abdalla:1981yt}. For the classical Lie algebra, we found that it was always possible to find a basis of a given mutually commuting subalgebra for which each independent $1d$ Lagrangian has a different, nonzero, finite order. For the quantum Lie algebra, this is not always the case.
\section{Discussion}
\label{sec:disc}
In this paper, we initiated a systematic approach to finding all possible mutually commuting symmetry algebras of putative $2d$ integrable models which can be obtained as a deformation of a field theory exhibiting left(right)-moving or (anti)-holomorphic currents. We saw how the Lie algebra of symmetries implied by such a current is best described on the vector space of $1d$ Lagrangian functionals $\ell(\{j_i(x^-)\},x^-)$ of the current $j(x^-)$. By having a full understanding of these Lie algebras, it is possible to find several differential equations that the generators $\ell(\{j_i(x^-)\},x^-)$ must satisfy so that they commute.

\sm

At least in the case of a single scalar, we were able to find all order 1 and 2 $1d$ Lagrangians which commute, with the order 1 Lagrangian taking the form
\begin{align}
\label{eq:discell1}
    \ell_1(j,j_1) =\sqrt{\Big(\frac{cj_1}{\sqrt{a(j)}}+b(j)\Big)^2+a(j)}+d(j) 
\end{align}
where $d'''(j)=0$ and $c$ a constant, and $a(j)$ and $b(j)$ satisfy
\begin{align}
 \frac{\partial^5}{\partial j^5}\Big(a(j)+b(j)^2\Big)=0
\end{align}
Any infinite dimensional mutually commuting subalgebra defining an integrable model containing an order 1 and 2 $1d$ Lagrangian must be within this parameter space of Lagrangians.

\sm

Furthermore, we were able to find all order 2 and 3 $1d$ Lagrangians which commute, and do not also commute with an order 1 Lagrangian. In this case, the order 2 Lagrangian takes the form
\begin{align}
\label{eq:discell2}
    \ell_2(j,j_1,j_2) = \big(a j_2+b(j,j_1)\big)^{1/3}+c(j)
\end{align}
where $a$ is a constant, $c'''(j)=0$, and $b(j,j_1)$ satisfies (\ref{eq:bjj1}), (\ref{eq:Bredef}), (\ref{eq:DE1}), (\ref{eq:DE2}), and (\ref{eq:A}). All known $2d$ integrable models involving a single scalar have symmetry algebras which contain order 1 or 2 Lagrangians which can be obtained from either (\ref{eq:discell1}) or (\ref{eq:discell2}) by taking special limits of the free parameters.

\sm

There are of course, several open problems remaining. The most pressing is what subspace of the parameter space of mutually commuting subalgebras mentioned above, which are at least two dimensional, are actually infinite dimensional? At least in the cases studied in \autoref{ssec:scale}, whenever you find two Lagrangians which commute, they automatically commute with several other Lagrangians. It may be possible to use one of the standard techniques for proving the existence of infinitely many commuting charges, such as finding a useful Lax pair/connection, B\"{a}cklund transformation, or recursion operator, although this is not guaranteed. The new sequences of mutually commuting Lagrangians found in \autoref{ssec:scale} for instance do not appear to lend themselves easily to these techniques, although the Korteweg–de Vries type models they define have special properties such as exact wave solutions. Typically the techniques for demonstrating integrability rely on the specifics of the dynamics in the model, and not the symmetries in their own right. It is desirable instead to develop techniques to prove whether a given mutually commuting subalgebra is infinite dimensional by studying only the Lie algebra, and not the putative integrable model. In the future, we intend on elaborating further on this for the $[j]=0,-1,-2$ sequences.

\sm

To map out all possible infinite dimensional mutually commuting subalgebras of (\ref{eq:scalarlie}), it amounts to solving the equation $\mathcal{E}(\partial_-\mathcal{E}(\ell_n)\mathcal{E}(\ell_m))=0$ with $m>n$, demanding that $\ell_n$ is not a quadratic functional of $j_n$. If there is a solution to this equation, $\ell_n$ would define the beginning of a mutually commuting subalgebra, having the lowest nonzero order in the sequence. Pushing this program beyond the examples in this paper is computationally challenging, but worth pursuing. 

\sm

In this paper, we only studied in detail the Lie algebra associated with a single scalar. In the future, it will be important to extend this analysis to other theories with a left-moving current. The Lie algebra associated with a collection of scalars must have mutually commuting subalgebras associated with Toda field theory, and there may be interesting generalizations of the subalgebras discovered in \autoref{sssec:j=0}, \autoref{sssec:j-1}, and \autoref{sssec:j-2}. The Lie algebras associated with a collection of fermions and Wess–Zumino–Witten models too deserve further study.

\sm

After quantization, the Lie algebras are deformed dramatically. Looking at the algebra for a single scalar (\ref{eq:quantumscalarLie}), the simple structure of the classical Lie algebra is obscured. In the future we would like to determine if it is possible to write (\ref{eq:quantumscalarLie}) in terms of Euler operators or higher Euler operators, so that it may be possible to leverage the techniques used in the classical case \cite{olver1993applications}. We suspect this will also help in understanding the universal structure of the quantum Lie algebra, perhaps extending this to theories with any central charge $c$ as in \cite{Sasaki:1987mm, ZAMOLODCHIKOV1989641}. However, the Lie algebras in \autoref{sec:scalars}, \autoref{sec:fermions}, and \autoref{sec:WZW} are quite different in nature, indicating that the quantum Lie algebras require more input than the central charge to define, which is contrary to the conjecture in \cite{Sasaki:1987mm}. 

\subsection*{Acknowledgements}
The author would like to thank Sriram Bharadwaj, Eric D'Hoker, Stephen Ebert, Yan-Yan Li, and Zhengdi Sun for invaluable discussions during the preparation of this work. This research was supported in part by the Mani L. Bhaumik Institute for Theoretical Physics.

\appendix
\section{Other examples}
\label{app:other}
In this appendix, we present other examples of Lie algebras from $2d$ classical field theories generated by left-moving currents, including a collection of free massless left-moving fermions, and Wess–Zumino–Witten models. Although we will not pursue a detailed analysis of mutually commuting subalgebras of these examples in this paper, this appendix serves to demonstrate the broad scope this method has for finding integrable deformations of theories with a left-moving current.
\subsection{$2d$ free massless fermions}
\label{sec:fermions}
The classical theory of $N$ free massless left-moving anti-commuting fermions $\psi^a$ has the action
\begin{align}
\label{eq:fermaction}
    S=\int d^2x\,\psi^a\partial_+\psi^a
\end{align}
The Euler–Lagrange equation $\partial_+\psi^a=0$ demands that $\psi^a(x^+,x^-)=\psi^a(x^-)$ is left-moving. This theory therefore has $N$ independent left-moving currents $j^a(x^-)=\psi^a$. The fundamental Poisson bracket for this theory is
\begin{align}
    \{j^a(x^-),j^b(y^-)\} = \frac{1}{2}\delta^{ab}\delta(x^--y^-)
\end{align}
through which the Lie algebra of symmetries may be derived.
\subsubsection{Lie algebra}
\label{ssec:fermlie}
The action (\ref{eq:fermaction}) is invariant under the infinitesimal transformation
\begin{align}
\label{eq:fermtrans}
    &\psi^a\to\psi^a + \epsilon\delta_{\ell}\psi^a \\
\label{eq:fermtrans1}
    &\delta_\ell\psi^a = \frac{1}{2}\mathcal{E}^a(\ell) = \frac{1}{2}\sum_{k=0}^{\infty}(-1)^k\partial_-^k\frac{\partial}{\partial j^a_k(x^-)}\ell(\{j^b_i(x^-)\},x^-)
\end{align}
for any functional $\ell(\{j^b_i(x^-)\},x^-)$, and $\epsilon$ is an infinitesimal parameter which is commuting or anti-commuting depending on $\ell$ to ensure that $\epsilon\delta_\ell\psi^a$ is anti-commuting. Under this transformation, the action becomes
\begin{align}
    \delta_\ell S&=(-1)^{|\ell|}\int d^2x\bigg(\partial_+\Big(\frac{1}{2}\psi^a\mathcal{E}^a(\ell)-\ell(\{j^b_i(x^-)\},x^-)\Big) \nonumber \\
     &\hspace{3cm}-\partial_-\Big(\sum_{k=0}^{\infty}\sum_{j=0}^{k-1}(-1)^{k-j}\partial_+\partial_-^{j}\psi^a\partial_-^{k-j-1}\frac{\partial}{\partial j_k^a(x^-)}\ell(\{j^b_i(x^-)\},x^-)\Big)\bigg) \nonumber \\
     &=0
\end{align}
where $|\ell|=0,1$ if $\ell$ is commuting or anti-commuting, respectively. The Noether current is $j_-=\ell(\{j^a_i(x^-)\},x^-)$ and $j_+=0$. 

\sm

The vector space of all transformations of the form (\ref{eq:fermtrans}) generates a graded Lie algebra, together with the commutator $[\epsilon\delta_\ell,\epsilon'\delta_{\ell'}]\psi^a$, which by direct computation is
\begin{align}
\label{eq:fermdirect}
    [\epsilon\delta_\ell,\epsilon'\delta_{\ell'}]\psi^a = \frac{1}{4}\sum_{k=0}^{\infty}\Big(\partial_-^k(\epsilon\,\mathcal{E}^b(\ell))\frac{\partial}{\partial j_k^b(x^-)}(\epsilon'\,\mathcal{E}^a(\ell'))-\partial_-^k(\epsilon'\,\mathcal{E}^b(\ell'))\frac{\partial}{\partial j_k^b(x^-)}(\epsilon\,\mathcal{E}^a(\ell))\Big)
\end{align}
Using the product and composition rules (\ref{eq:eulerprod}) and (\ref{eq:eulercomp}), the right hand side of this equation may be written in the form of a new transformation $\frac{1}{2}\epsilon\epsilon'\mathcal{E}^a(\ell'')$. Stripping off the factor $\epsilon\epsilon'(-1)^{|\ell||\ell'|}$ to the left, (\ref{eq:fermdirect}) becomes a graded commutation relation $[\epsilon\delta_\ell,\epsilon'\delta_{\ell'}]=\epsilon\epsilon'(-1)^{|\ell||\ell'|}[\delta_\ell,\delta_{\ell'}\}$, where $[A,B\}\equiv AB-(-1)^{|A||B|}BA$
\begin{align}
\label{eq:fermlie}
    [\delta_\ell,\delta_{\ell'}\}\psi^a =\delta_{\ell''}\psi^a = \frac{1}{2}\mathcal{E}^a(\ell''),\qquad \ell''=\frac{1}{4}(-1)^{|\ell|}\{\mathcal{E}^b(\ell),\mathcal{E}^b(\ell')]
\end{align}
where we have defined a graded \textit{anti}-commutator $\{A,B]\equiv AB+(-1)^{|A||B|}BA$. Here $\ell''$ is written in a way which manifestly transforms like the graded commutator $[\delta_\ell,\delta_{\ell'}\}$ under $(\ell\leftrightarrow \ell')$, although it may be written more simply as $\frac{1}{2}(-1)^{|\ell|}\mathcal{E}^b(\ell)\mathcal{E}^b(\ell')$.

\sm

To show that (\ref{eq:fermdirect}) and (\ref{eq:fermlie}) are equivalent, we first write the right hand side (\ref{eq:fermdirect}) with the factor $\epsilon\epsilon'(-1)^{|\ell||\ell'|}$ stripped off to the left.
\begin{align}
    \frac{1}{4}\sum_{k=0}^{\infty}\Big(\partial_-^k\mathcal{E}^b(\ell)\frac{\partial}{\partial j_k^b(x^-)}\mathcal{E}^a(\ell')-(-1)^{|\ell||\ell'|}\partial_-^k\mathcal{E}^b(\ell')\frac{\partial}{\partial j_k^b(x^-)}\mathcal{E}^a(\ell)\Big)
\end{align}
Next, we compute $\frac{1}{2}\mathcal{E}^a(\ell'')$
\begin{align}
    \frac{1}{2}\mathcal{E}^a(\ell'') &= \frac{1}{4}(-1)^{|\ell|}\mathcal{E}^a(\mathcal{E}^b(\ell)\mathcal{E}^b(\ell'))\nonumber \\
    &=\frac{1}{4}(-1)^{|\ell|}\sum_{k=0}^{\infty}(-1)^k\Big(\mathcal{E}^a_k(\mathcal{E}^b(\ell))\partial_-^k\mathcal{E}^b(\ell')-(-1)^{|\ell|}\partial_-^k\mathcal{E}^b(\ell)\mathcal{E}^a_k(\mathcal{E}^b(\ell'))\Big)\nonumber \\
    &=\frac{1}{4}(-1)^{|\ell|}\sum_{k=0}^{\infty}\Big(-\frac{\partial}{\partial j_k^b(x^-)}\mathcal{E}^a(\ell)\partial_-^k\mathcal{E}^b(\ell')+(-1)^{|\ell|}\partial_-^k\mathcal{E}^b(\ell)\frac{\partial}{\partial j_k^b(x^-)}\mathcal{E}^a(\ell')\Big) \nonumber \\
    &=\frac{1}{4}\sum_{k=0}^{\infty}\Big(\partial_-^k\mathcal{E}^b(\ell)\frac{\partial}{\partial j_k^b(x^-)}\mathcal{E}^a(\ell')-(-1)^{|\ell||\ell'|}\partial_-^k\mathcal{E}^b(\ell')\frac{\partial}{\partial j_k^b(x^-)}\mathcal{E}^a(\ell)\Big)
\end{align}
where the product rule (\ref{eq:eulerprod}) was used in the second line, and the composition rule (\ref{eq:eulercomp}) was used in the third line, both taking into account the anti-commuting nature of $\psi^a$. This proves the equivalence of (\ref{eq:fermdirect}) and (\ref{eq:fermlie}).
\subsubsection{Subalgebras}
\label{ssec:fermsub}
The subalgebras of the graded Lie algebra (\ref{eq:fermlie}) of symmetry transformations of the action (\ref{eq:fermaction}) afford a similar description as in \autoref{ssec:scalarsub}. Here we describe some of this structure.

\sm

Because $j^a(x^-)=\psi^a$ is an anti-commuting variable, $\ell(\{j^a_i(x^-)\},x^-)$ must be a polynomial functional of $j^a(x^-)$ and its derivatives. If $\ell$ and $\ell'$ are degree $n$ and $m$ polynomial functionals respectively, $\ell''$ in (\ref{eq:fermlie}) will be at most a degree $n+m-2$ polynomial. Lagrangians $\ell$ of degree 2 and less therefore form a subalgebra. This is the linear subalgebra.

\sm

The space of $1d$ Lagrangians $\ell$ which commute with the energy-momentum tensor $T=j^a\partial_-j^a$ are again the ones with a conserved Hamiltonian, because
\begin{align}
    \frac{1}{2}(-1)^{|T|}\mathcal{E}^b(T)\mathcal{E}^b(\ell)=\partial_-j^b\mathcal{E}^b(\ell)&=\sum_{k=0}^{\infty}(-1)^k\partial_-j^b\partial_-^k\frac{\partial}{\partial j_k^b(x^-)}\ell(\{j_i^a(x^-)\},x^-) \nonumber \\
    &\sim \sum_{k=0}^{\infty}\partial_-j_k^b\frac{\partial}{\partial j_k^b(x^-)}\ell(\{j_i^a(x^-)\},x^-) \nonumber \\
    & = \frac{d}{dx^-}\ell-\frac{\partial}{\partial x^-}\ell \nonumber \\
    &\sim -\frac{\partial}{\partial x^-}\ell(\{j_i^a(x^-)\},x^-)
\end{align}
and so for $\ell(\{j_i^a(x^-)\},x^-)$ to commute with $T$, $\ell=\ell(\{j_i^a(x^-)\})$ must have no explicit $x^-$ dependence. This is a subalgebra.

\sm

The Witt algebra is a subalgebra, with Lagrangian generators $l_n=(x^-)^{n+1}T$. Indeed, the commutator between the two generators $l_n$ and $l_m$ is
\begin{align}
    \frac{1}{2}\mathcal{E}(l_n)\mathcal{E}(l_m)&=\frac{1}{2}\big((n+1)(x^-)^nj^a+2(x^-)^{n+1}\partial_-j^a\big)\big((m+1)(x^-)^mj^a+2(x^-)^{m+1}\partial_-j^a\big)\nonumber \\
    &= (n-m)(x^-)^{n+m+1}T=(n-m)l_{n+m}
\end{align}

\sm

As before, due to the $O(N)$ invariance of the action (\ref{eq:fermaction}), if $\ell$ and $\ell'$ are $O(N)$ invariant, then the commutator (\ref{eq:fermlie}) is also $O(N)$ invariant, and so the space of $O(N)$ invariant $1d$ Lagrangians forms a subalgebra. In this case, only $1d$ Lagrangians with $|\ell|=0$ can be $O(N)$ invariant.

\sm

If two $1d$ Lagrangians $\ell$ and $\ell'$ commute, then $\ell$ is invariant under the infinitesimal transformation (and vice versa)
\begin{align}
    \delta_{\ell'} j^a = \frac{1}{2}\mathcal{E}^a(\ell')
\end{align}
which is just the transformation (\ref{eq:fermtrans1}), since $j^a=\psi^a$. Indeed, $\ell$ transforms as
\begin{align}
    \delta_{\ell'}\ell(\{j_i^a(x^-)\},x^-)&=\frac{1}{2}\sum_{k=0}^{\infty}\partial_-^{k}(\mathcal{E}^b(\ell'))\frac{\partial}{\partial j^b_k(x^-)}\ell(\{j_i^a(x^-)\},x^-) \nonumber \\
    &\sim \frac{1}{2}\sum_{k=0}^{\infty}(-1)^k\mathcal{E}^b(\ell')\partial_-^{k}\frac{\partial}{\partial j^b_k(x^-)}\ell(\{j_i^a(x^-)\},x^-) \nonumber \\
    &=\frac{1}{2}\mathcal{E}^b(\ell')\mathcal{E}^b(\ell)\sim 0
\end{align}
where the last line is equivalent to zero because $\ell$ and $\ell'$ commute in the sense of (\ref{eq:fermlie}).

\sm

Because of this, if there exists an infinite dimensional mutually commuting subalgebra, with a countably infinite set of Lagrangian generators $\ell_n$ with $n=1,2,\dots,\infty$, then there exists infinitely many integrable deformations of the classical $N$ free massless left-moving fermion action (\ref{eq:fermaction}) of the form
\begin{align}
    S=\int d^2x\Big(\psi^a\partial_+\psi^a+\lambda\,\ell_n(\{j_i^b\},x^-)\Big)
\end{align}
This action is invariant under the infinitesimal transformation $\delta_{\ell_m}\psi^a=\frac{1}{2}\mathcal{E}^a(\ell_m)$ for all $n,m$ and finite $\lambda$.
\subsection{Wess–Zumino–Witten models}
\label{sec:WZW}
The classical Wess–Zumino–Witten model \cite{Witten:1983ar} on a Lie group $G$ and its associated Lie algebra $\mathfrak{g}$ as the action
\begin{align}
\label{eq:WZWaction}
    S_{\text{WZW}} =- \frac{|k|}{4\pi}\int d^2x\,\text{tr}(g^{-1}\partial_+gg^{-1}\partial_-g) + k\Gamma
\end{align}
with $g\in G$. The trace is normalized such that $\text{tr}(t^it^j)=\frac{1}{2}\delta^{ij}$ for two generators of the Lie algebra in the fundamental representation $(t^i)^{ab}$. $\Gamma$ is the Wess–Zumino term
\begin{align}
    \Gamma = \frac{1}{12\pi}\int d^3y\epsilon_{\alpha\beta\gamma}\text{tr}(g^{-1}\partial^{\alpha}gg^{-1}\partial^{\beta}gg^{-1}\partial^{\gamma}g)
\end{align}

\sm

Classically, there is no constraint on $k$, but it may be constrained after quantization. For instance, $k$ must be an integer for a compact Lie group. For the rest of this section, we will for simplicity assume $k>0$. In this case, this theory admits a left-moving $\mathfrak{g}$ valued current $j^{ab}(x^-)=(\partial_-gg^{-1})^{ab}$ (and a right-moving $\mathfrak{g}$ valued current $(g^{-1}\partial_+g)^{ab}$). The fundamental Poisson bracket for the left-moving sector is (up to a normalization)
\begin{align}
    \{j^{ab}(x^-),j^{cd}(y^-)\} = \frac{1}{2}\big(j^{ad}(x^-)\delta^{cb}-\delta^{ad}j^{cb}(x^-)\big)\delta(x^--y^-)-\frac{1}{2}\delta^{ad}\delta^{cb}\partial_{x^-}\delta(x^--y^-)
\end{align}
through which the Lie algebra of symmetries, or current algebra, may be derived \cite{Knizhnik:1984nr}. We will at times suppress the $\mathfrak{g}$ indices $ab$ for notational simplicity.
\subsubsection{Lie algebra}
\label{ssec:WZWlie}
This action is invariant under the infinitesimal transformation
\begin{align}
\label{eq:WZWtrans}
& g\to g+\epsilon\delta_\ell g \nonumber \\
    &\delta_\ell g=\frac{1}{2}\mathcal{E}(\ell)g,\quad \mathcal{E}^{ab}(\ell)=\sum_{k=0}^{\infty}(-1)^k\partial_-^k\frac{\partial}{\partial j^{ba}_k(x^-)}\ell(\{j_i(x^-)\},x^-)
\end{align}
for any functional $\ell(\{j_i(x^-)\},x^-)$, and $\epsilon$ is an infinitesimal parameter. Under this transformation, the action becomes
\begin{align}
    \delta_\ell S&=-\frac{k}{4\pi}\int d^2x\,\text{tr}\bigg(\partial_+\Big(\frac{1}{2}\mathcal{E}(\ell)\partial_-gg^{-1}-\ell(\{j_i(x^-)\},x^-)\Big) \nonumber\\
    &\hspace{2cm}+\partial_-\Big(\frac{1}{2}\mathcal{E}(\ell)\partial_+gg^{-1}-\sum_{k=0}^{\infty}\sum_{j=0}^{k-1}(-1)^{k-j}\partial_-^{k-j-1}\frac{\partial}{\partial j_k(x^-)}\ell\,\partial_+\partial_-^j(\partial_-gg^{-1})\Big)\bigg) \nonumber \\
    &=0
\end{align}
and so the Noether current is $j_-=\ell(\{j_i(x^-)\},x^-)$ and $j_+=0$. As before, the symmetry transformations (\ref{eq:WZWtrans}) associated with two Lagrangians $\ell$ and $\ell'$ do not in general commute. By direct computation, the commutator $[\delta_\ell,\delta_{\ell'}]g$ is
\begin{align}
\label{eq:WZWdirect}
    &([\delta_\ell,\delta_{\ell'}]g)^{ab} = M^{ac}g^{cb} \\
    & M^{ab}= \nonumber \\
    &\frac{1}{4}\bigg(\sum_{k=0}^{\infty}\Big(\partial_-^kD_-\mathcal{E}^{cd}(\ell)\frac{\partial}{\partial j_k^{cd}(x^-)}\mathcal{E}^{ab}(\ell')-\partial_-^kD_-\mathcal{E}^{cd}(\ell')\frac{\partial}{\partial j_k^{cd}(x^-)}\mathcal{E}^{ab}(\ell)\Big)-[\mathcal{E}(\ell),\mathcal{E}(\ell')]^{ab}\bigg)
\end{align}
where $D_-A=\partial_-A-[j,A]$ is the covariant derivative with $\mathfrak{g}$ valued connection $j$ for $A$ in the adjoint representation. The right hand side of this equation may be written in the form of a new transformation $\frac{1}{2}\mathcal{E}(\ell'')g$, where $\ell''$ is a new Lagrangian
\begin{align}
\label{eq:WZWLie}
    [\delta_\ell,\delta_{\ell'}]g = \delta_{\ell''}g=\frac{1}{2}\mathcal{E}(\ell'')g,\qquad \ell''=\frac{1}{4}\text{tr}\Big(D_-\mathcal{E}(\ell)\mathcal{E}(\ell')-D_-\mathcal{E}(\ell')\mathcal{E}(\ell)\Big)
\end{align}
where $\ell''$ is written in a manifestly anti-symmetric in $(\ell\leftrightarrow \ell')$ way, although we may write it more simply as $\ell''=\frac{1}{2}\text{tr}(D_-\mathcal{E}(\ell)\mathcal{E}(\ell'))$. To show that (\ref{eq:WZWdirect}) and (\ref{eq:WZWLie}) are equivalent, we compute $\frac{1}{2}\mathcal{E}^{ab}(\ell'')$
\begin{align}
\label{eq:WZWequiv1}
    &\frac{1}{2}\mathcal{E}^{ab}(\ell'')=\frac{1}{4}\mathcal{E}^{ab}(\text{tr}(D_-\mathcal{E}(\ell)\mathcal{E}(\ell')))\nonumber \\
    &=\frac{1}{4}\text{tr}\sum_{k=0}^{\infty}(-1)^k\Big(\mathcal{E}^{ab}_k(D_-\mathcal{E}(\ell))\partial_-^k\mathcal{E}(\ell')+\partial_-^kD_-\mathcal{E}(\ell)\mathcal{E}^{ab}_k(\mathcal{E}(\ell'))\Big) \nonumber \\
    &=\frac{1}{4}\text{tr}\sum_{k=0}^{\infty}(-1)^k\Big(\partial_-^{k}D_-\mathcal{E}(\ell)\mathcal{E}_k^{ab}(\mathcal{E}(\ell'))-\partial_-^{k+1}\mathcal{E}(\ell')\mathcal{E}_k^{ab}(\mathcal{E}(\ell))-\mathcal{E}^{ab}_k([j,\mathcal{E}(\ell)])\partial_-^k\mathcal{E}(\ell')\Big)
\end{align}
where the product rule (\ref{eq:eulerprod}) was used in the second line. The third line is obtained after writing $\mathcal{E}_k^{ab}(D_-\mathcal{E}(\ell))=\mathcal{E}_{k-1}^{ab}(\mathcal{E}(\ell))-\mathcal{E}_k^{ab}([j,\mathcal{E}(\ell)])$. The last term on the third line can be simplified and written as
\begin{align}
   \text{tr}\sum_{k=0}^{\infty}(-1)^k\Big(\mathcal{E}^{ab}_k([j,\mathcal{E}(\ell)])\partial_-^k\mathcal{E}(\ell')\Big) = [\mathcal{E}(\ell),\mathcal{E}(\ell')]^{ab}-\text{tr}\sum_{k=0}^{\infty}(-1)^k\Big(\partial_-^k[j,\mathcal{E}(\ell')]\mathcal{E}_k^{ab}(\mathcal{E}(\ell))\Big)
\end{align}
Substituting this into (\ref{eq:WZWequiv1}), we get
\begin{align}
   & \frac{1}{2}\mathcal{E}^{ab}(\ell'')= \nonumber \\
    &=\frac{1}{4}\bigg(\text{tr}\sum_{k=0}^{\infty}(-1)^k\Big(\partial_-^{k}D_-\mathcal{E}(\ell)\mathcal{E}_k^{ab}(\mathcal{E}(\ell'))-\partial_-^{k}D_-\mathcal{E}(\ell')\mathcal{E}_k^{ab}(\mathcal{E}(\ell))\Big) -[\mathcal{E}(\ell),\mathcal{E}(\ell')]^{ab}\bigg)\nonumber \\
    & = \frac{1}{4}\bigg(\sum_{k=0}^{\infty}\Big(\partial_-^kD_-\mathcal{E}^{cd}(\ell)\frac{\partial}{\partial j_k^{cd}(x^-)}\mathcal{E}^{ab}(\ell')-\partial_-^kD_-\mathcal{E}^{cd}(\ell')\frac{\partial}{\partial j_k^{cd}(x^-)}\mathcal{E}^{ab}(\ell)\Big)-[\mathcal{E}(\ell),\mathcal{E}(\ell')]^{ab}\bigg)
\end{align}
where the composition rule (\ref{eq:eulercomp}) was used in the third line. This proves the equivalence of (\ref{eq:WZWdirect}) and (\ref{eq:WZWLie}).
\subsubsection{Subalgebras}
\label{ssec:WZWsub}
The subalgebras of the Lie algebra (\ref{eq:WZWLie}) of symmetry transformations of the action (\ref{eq:WZWaction}) again afford a similar description as in \autoref{ssec:scalarsub} and \autoref{ssec:fermsub}.

\sm

If $\ell$ and $\ell'$ are polynomial functionals of degree $n$ and $m$ respectively of $j(x^-)$ and its derivatives, $\ell''$ in (\ref{eq:WZWLie}) will be at most a degree $n+m-1$ polynomial, instead of $n+m-2$, because of the appearance of $j$ in the derivative $D_-=\partial_--[j,\,\cdot\,]$. In this case, Lagrangians $\ell$ of degree 1 and less therefore form a subalgebra.

\sm

The space of $1d$ Lagrangians $\ell$ which commute with the energy-momentum tensor $T=\text{tr}(j^2)$ are again the ones with a conserved Hamiltonian, because
\begin{align}
    \frac{1}{2}\text{tr}(D_-\mathcal{E}(T)\mathcal{E}(\ell)) =\text{tr}(\partial_-j\mathcal{E}(\ell)) &= \sum_{k=0}^{\infty}(-1)^k\partial_-j^{ba}\partial_-^k\frac{\partial}{\partial j_k^{ba}(x^-)}\ell(\{j_i(x^-)\},x^-) \nonumber \\
    &\sim \sum_{k=0}^{\infty}\partial_-j_k^{ba}\frac{\partial}{\partial j_k^{ba}(x^-)}\ell(\{j_i(x^-)\},x^-)\nonumber\\
    &= \frac{d}{dx^-}\ell-\frac{\partial}{\partial x^-}\ell\nonumber\\
    &\sim-\frac{\partial}{\partial x^-}\ell(\{j_i(x^-)\},x^-)
\end{align}
and so for $\ell(\{j_i(x^-)\},x^-)$ to commute with $T$, $\ell=\ell(\{j_i(x^-)\})$ must have no explicit $x^-$ dependence. This is a subalgebra.

\sm

The Witt algebra is a subalgebra, with Lagrangian generators $l_n=(x^-)^{n+1}T$. Indeed, the commutator between the two generators $l_n$ and $l_m$ is
\begin{align}
    \frac{1}{2}\text{tr}(D_-\mathcal{E}(l_n)\mathcal{E}(l_m))&=2(n+1)(x^-)^{n+m+1}T + (x^-)^{n+m+2}\partial_-T\nonumber \\
    &\sim (n-m)(x^-)^{n+m+1}T=(n-m)l_{n+m}
\end{align}

\sm

If a $1d$ Lagrangian $\ell(\{j_i(x^-)\},x^-)$ is invariant under global $g\in G$ transformations $j\to gjg^{-1}$, then its Euler–Lagrange equation $\mathcal{E}(\ell)$ transforms covariantly $\mathcal{E}(\ell)\to g\mathcal{E}(\ell)g^{-1}$. Looking at the form of the commutator (\ref{eq:WZWLie}), if both $\ell$ and $\ell'$ are $G$ invariant, $\ell''$ is as well. The space of $G$ invariant $1d$ Lagrangians therefore forms a subalgebra as well.

\sm

If two $1d$ Lagrangians $\ell$ and $\ell'$ commute, then $\ell$ is invariant under the infinitesimal transformation (and vice versa)
\begin{align}
    \delta_{\ell'}j =\frac{1}{2}D_-\mathcal{E}(\ell') 
\end{align}
which is just the transformation (\ref{eq:WZWtrans}) acting on $j=\partial_-gg^{-1}$. Performing this transformation on $\ell(\{j_i(x^-)\},x^-)$, we get
\begin{align}
    \delta_{\ell'}\ell(\{j_i(x^-)\},x^-) &= \frac{1}{2}\sum_{k=0}^{\infty}\partial_-^k(D_-\mathcal{E}(\ell'))^{ab}\frac{\partial}{\partial j_k^{ab}(x^-)}\ell(\{j_i(x^-)\},x^-) \nonumber \\
    &\sim \frac{1}{2}\sum_{k=0}^{\infty}(-1)^k(D_-\mathcal{E}(\ell'))^{ab}\partial_-^k\frac{\partial}{\partial j_k^{ab}(x^-)}\ell(\{j_i(x^-)\},x^-) \nonumber\\
    & = \frac{1}{2}\text{tr}(D_-\mathcal{E}(\ell')\mathcal{E}(\ell))\sim 0
\end{align}
where the last line is equivalent to zero because $\ell$ and $\ell'$ commute in the sense of (\ref{eq:WZWLie}).

\sm

Because of this, if there exists an infinite dimensional mutually commuting subalgebra, with a countably infinite set of Lagrangian generators $\ell_n$ with $n=1,2,\dots,\infty$, then there exists infinitely many integrable deformations of the classical Wess–Zumino–Witten model action (\ref{eq:WZWaction}) of the form
\begin{align}
    S = S_{\text{WZW}} + \lambda\int d^2x\,\ell_n(\{j_i\},x^-)
\end{align}
This action is invariant under the infinitesimal transformation $\delta_{\ell_m}g=\frac{1}{2}\mathcal{E}(\ell_m)g$ for all $n,m$ and finite $\lambda$.

\bibliography{Nonlocal}
\end{document}